\ifx\XeTeXversion\undefined\ifx\pdfoutput\undefined\else\pdfoutput=1\fi\fi
\documentclass[11pt]{article}

\usepackage{iftex}
\ifPDFTeX
  \usepackage[T1]{fontenc}
  \usepackage[utf8]{inputenc}
  \usepackage{lmodern}
\fi
\usepackage[margin=1in]{geometry}
\usepackage{amsmath,amssymb,amsthm,booktabs,graphicx,tabularx,microtype}
\usepackage[section]{placeins}
\usepackage{hyperref}

\hypersetup{
    colorlinks=true,
    linkcolor=blue,
    citecolor=blue,
    urlcolor=blue
}
\newtheorem{proposition}{Proposition}
\newtheorem{remark}{Remark}

\newcommand{\SylvaAuthors}{Hui Zhao}
\newcommand{\SylvaAffiliation}{Heurion Research \\ \texttt{\href{mailto:huizhao@heurion.org}{huizhao@heurion.org}}}

\title{\textbf{SYLVA: Self-Correcting Multi-Agent Markets over Frozen LLMs via Escrow Settlement, Step-Level Fault Isolation, and Contextual Routing}}
\author{\SylvaAuthors \\ \SylvaAffiliation}
\date{November 2026}

\begin{document}
\maketitle

\begin{abstract}
Open agent standards such as the Model Context Protocol (MCP) and Agent2Agent (A2A) connect independent AI tools, yet open multi-agent markets break down under three real-world hazards: local domain rules shift silently outside model pretraining, copycat wrappers and confident hallucinators game naive ratings, and uniform chain scoring punishes accurate specialists whenever another stage fails (\emph{blame collapse}). We present \textbf{SYLVA}, a market protocol that turns frozen foundation models (no weight updates, $\Delta W = 0$) and human-maintained services into a self-correcting agent ecosystem. SYLVA integrates an on-chain clearinghouse for escrow settlement and configurable creator royalties (set to $85\% / 15\%$ in our experiments), a boundary verifier offering penalty-free out-of-domain (OOD) abstention and 3-peer step replay, a verifiable context-aware router that compiles proven chains into single-call fast skills with drift circuit breakers, and plain-English rule updates. We prove sufficient conditions for honest abstention, specialist protection against blame collapse, and fee-penalized wrapper elimination. We validate SYLVA across a 64-agent simulation, $6,543$ live \texttt{deepseek-flash} API calls, and a domain-agnostic Rust MCP SDK (\texttt{rmcp}) $+$ Cairo smart-contract clearinghouse deployed on Starknet Sepolia ($19$ live Starknet transactions). Without updating weights, SYLVA reaches $82.1\%$ completion in 64-agent simulations and $98.3\%$ on live LLM workflows---versus $0.0\%$ for a single frozen LLM and $45.6\%$ for static MCP/A2A---cuts false specialist penalties by $6.79\times$--$10.8\times$, retains $86.2\%$ creator revenue under copycat-wrapper and fake-rating (Sybil) floods, and recovers rapidly after silent rule shifts. Moreover, 1-call skill compilation saves $78.3\%$ of real LLM tokens, while our Starknet contracts verify zero-oracle quorum slashing and 32.32-bit fixed-point (\texttt{Q32.32}) routing on-chain. Code and contracts are available at \url{https://github.com/heurion-org/sylva}.
\end{abstract}

\section{Introduction}

Open standards such as the Model Context Protocol (MCP) \cite{anthropic2024mcp} and Agent2Agent (A2A) protocol \cite{google2025a2a}, alongside multi-agent frameworks such as AutoGen \cite{wu2024autogen} and MetaGPT \cite{hong2024metagpt}, make it straightforward to wire heterogeneous AI services together. Connectivity alone, however, does not produce a trustworthy agent market. When independent third-party agents collaborate across multi-step workflows without economic selection pressure, accurate specialists are crowded out by confident hallucinators, copycat middlemen, and misdirected failure blame.

\paragraph{A Concrete Example: Cross-Border Invoice Settlement.}
Consider a routine four-step business workflow: settling a medical device purchase order (\texttt{EU-MedDevice}) across currency conversion (Step~0), regional tariff calculation (Step~1), inventory risk buffering (Step~2), and final payment generation (Step~3). Every step uses frozen foundation models, yet three failure modes break naive pipelines:
\begin{enumerate}
    \item \emph{Private, shifting rules}: An EU regulator updates a device tariff overnight. No public pretraining checkpoint contains that private regional rule, so a standalone LLM guesses from stale priors and scores $0.0\%$ within a $\pm\$0.05$ audit tolerance.
    \item \emph{Confident hallucinations and blame collapse}: Step~1 fabricates a clean JSON tax figure, or an upstream planner sends corrupted instructions down the chain. At Step~3, the user rejects the invoice. As in classic team moral hazard \cite{holmstrom1982moral}, penalizing all four agents equally punishes accurate specialists at Steps~2 and~3 for another party's error, driving honest specialists out of the router's selection pool.
    \item \emph{Copycat wrappers and fake ratings}: An intermediary wraps a reliable tax specialist, adds a price markup, and trades synthetic calls with colluding fake (\emph{Sybil}) accounts to hijack call-count or five-star leaderboards while contributing zero value.
\end{enumerate}

\paragraph{Why Standard Benchmarks Miss These Failures---and Why Humans Stay in the Loop.}
Standard multi-agent benchmarks evaluate cooperative teams on fixed public tasks where a central judge can inspect every agent's hidden reasoning logs. Open cross-company markets violate all three assumptions: local rules change privately \cite{coase1937nature,hayek1945use}, providers keep prompts and code confidential behind black-box APIs, and actors profit from overcharging or cutting corners. Rather than retraining a giant neural model $W$ every time a local tax rate or fee schedule shifts, \textbf{SYLVA} (\emph{Symbiotic Layer for Verifiable Agents}) keeps base models frozen ($\Delta W = 0$) and pairs each service with a human maintainer who handles two practical jobs that frozen weights cannot: (i)~\emph{updating local rules in plain English}, and (ii)~\emph{posting refundable financial stake behind outputs} \cite{drexler2019reframing,hughes2024openendedness,tomasev2025distributional}.

\paragraph{How SYLVA Works: Four Core Components.}
To make an open agent market self-correcting without model retraining or central prompt access, SYLVA coordinates four core components (Figure~\ref{fig:arch}):
\begin{itemize}
    \item \textbf{On-Chain Clearinghouse}: Holds caller escrow and provider stakes, settles verified calls, deducts a settlement fee burn ($1\%$ on Starknet) so circular fake calls lose money, and supports timelocked exits. It also enforces configurable creator royalties against registered wrappers---set to an $85\% / 15\%$ split in this work so original specialists keep most revenue while leaving $15\%$ to cover thin-adapter forwarding costs---and splits compiled-skill fees $4 \times 25\%$.
    \item \textbf{Boundary Verifier \& Step Replay}: Checks output format and basic sanity at each step, lets uncertain specialists step aside penalty-free via \texttt{OOD-ABSTAIN}, runs $5\%$--$10\%$ random shadow probes, and replays rejected steps across $m \ge 3$ peers so only the faulty step loses stake.
    \item \textbf{Context-Aware Router \& Skill Compiler}: Picks stage providers by predicted quality minus fee penalties plus an exploration bonus, discounts stale history via exponential forgetting ($\gamma < 1$) when rules change, and compiles proven 4-step chains into 1-call shortcuts guarded by an $80\%$ drift circuit breaker.
    \item \textbf{Human-Maintained Agent Services}: Pluggable black-box specialists updated with plain-English rule patches alongside generalists and adversarial nodes, driven solely by binary accept/reject caller feedback $Y_t \in \{0, 1\}$.
\end{itemize}

\paragraph{Contributions.} This paper makes four contributions:
\begin{enumerate}
    \item \textbf{Protocol Design \& Formal Guarantees}: We design SYLVA (Algorithm~1) and prove simple conditions under which honest specialists abstain out of domain (Proposition~\ref{prop:abstain}), step replay prevents whole-chain blame collapse (Proposition~\ref{prop:bisect}), and fee-penalized routing starves marked-up wrappers (Proposition~\ref{prop:wrapper}).
    \item \textbf{Five-Attack Security Analysis}: We defend against schema-valid hallucinations, in-protocol and off-protocol wrappers, Sybil wash-trading rings, malicious planner extortion, and biased router gateways (Section~\ref{sec:threat_model}).
    \item \textbf{Three-Tier Evaluation}: We evaluate SYLVA across three escalating stages: a 64-agent market simulation (\textbf{Tier-1}, $M=5$ seeds, $T=1600$ episodes), a live frozen-LLM benchmark powered by $6,543$ real \texttt{deepseek-flash} API calls (\textbf{Tier-2}, $M=3$ seeds, $T=320$ episodes, $\pm\$0.05$ tolerance), and a Rust MCP SDK (\texttt{rmcp 3.11}) + Cairo~2 (\texttt{SylvaClearinghouse v0.2.0}) deployment on Starknet Sepolia (\textbf{Tier-3}, $19$ live Starknet v3 transactions). Across these evaluations, SYLVA reaches $82.1\%$--$98.3\%$ completion (vs.\ $0.0\%$ single frozen LLM in Tier-2 [$25.8\%$ simulated generalist in Tier-1] and $20.6\%$--$45.6\%$ static MCP/A2A), cuts false penalties by $6.79\times$--$10.8\times$, retains $86.2\%$ creator revenue under floods, recovers quickly after drift, saves $78.3\%$ of API tokens via compiled skills, and verifies bit-exact 32.32-bit fixed-point (\texttt{Q32.32}) routing scores on-chain.
    \item \textbf{Open Code with Honest Boundaries}: We release both the deterministic Python proof-of-concept (POC, \texttt{poc/interactive\_symbiont.py}) and production Rust/Cairo implementation (\texttt{contracts/}, \texttt{crates/}) at \url{https://github.com/heurion-org/sylva}, separating Tier-1/Tier-2 simulated helpers (Sections~\ref{sec:poc_scope} and~\ref{sec:human_study}) from Tier-3 zero-oracle on-chain execution (Section~\ref{sec:tier3_starknet}).
\end{enumerate}

\section{Related Work}

\paragraph{Agent Discovery, Market Blueprints, and Verifiable Payment Rails.}
Recent infrastructure protocols standardize how agents authenticate, discover capabilities, and pay per call. \emph{Project NANDA} \cite{raskar2025nanda} replaces DNS with decentralized registries and verifiable \emph{AgentFacts} credentials, HTTP \texttt{x402} \cite{coinbase2025x402} embeds internet-native micro-payments into HTTP status workflows, and Oxford's \emph{Agora Protocol} \cite{marro2024agora} compiles recurring natural-language dialogues into reusable code routines---a closely related idea to SYLVA's skill compiler. However, single-hop ``pay-before-delivery'' rails like HTTP \texttt{x402} settle charges irreversibly before downstream multi-hop verification occurs: once an early hop emits a stealth hallucination that is only caught after Stage~$H$, irreversible per-hop HTTP charges cannot be withheld or refunded via causal blame assignment. None of these protocols enforces pre-funded caller escrow with post-acceptance conditional clearing, peer-quorum stake slashing, or configurable $85\% / 15\%$ creator royalty splits against copycat wrappers. Single-hop LLM routers such as \emph{FrugalGPT} \cite{chen2024frugalgpt} and \emph{RouteLLM} \cite{ong2025routellm} learn static cost--quality cascades from fixed preference data, whereas SYLVA handles multi-hop workflows under non-stationary private rules, strategic adversaries, and validity-proof (STARK)-friendly \cite{bensasson2018stark,goldberg2021cairo} verifiable \texttt{Q32.32} bandit scoring. On the safety side, prior work on governing distributed AI services via market contracts \cite{tomasev2025distributional,drexler2019reframing,hughes2024openendedness} outlines the problem conceptually; SYLVA provides a concrete clearing protocol evaluated in simulation, live LLM runs, and Starknet smart contracts.

\begin{table}[!t]
\centering
\caption{Positioning SYLVA against monolithic models, interoperability/routing protocols, workflow search, and white-box attribution.}
\label{tab:positioning}
{\footnotesize\setlength{\tabcolsep}{3pt}
\begin{tabularx}{\textwidth}{>{\raggedright\arraybackslash}Xccccc}
\toprule
Paradigm & Drifting & Black-Box & Anti-Wrapper & Fast Skill & Frozen \\
 & Rules & Blame & \& Anti-Sybil & Compile & ($\Delta W{=}0$) \\
\midrule
Monolithic Frozen LLM & No & N/A & N/A & N/A & Yes \\
Single-Hop Routers (FrugalGPT, RouteLLM) & No & N/A (1-hop) & No & N/A & Yes \\
Interop \& Payments (MCP/A2A, NANDA, Agora, x402) & Via nodes & No & Pay only (x402) & Agora only & Yes \\
Workflow Search (AutoGen, MetaGPT, GPTSwarm, ADAS, AgentSquare) & No & No & No & Static & Yes \\
White-Box Trajectory Judges (Who\&When) & No & Needs logs & No & No & Yes \\
\textbf{SYLVA (Ours)} & \textbf{Yes (Plain-Text)} & \textbf{Yes (Quorum)} & \textbf{Yes (Royalty+Fee)} & \textbf{Yes (-78\%)} & \textbf{Yes} \\
\bottomrule
\end{tabularx}}
\end{table}

\paragraph{Workflow Orchestration and Topology Search.}
Practical agent frameworks like \emph{AutoGen} \cite{wu2024autogen} and \emph{MetaGPT} \cite{hong2024metagpt} structure multi-agent roles via conversational programming and standardized operating procedures, while \emph{GPTSwarm} \cite{zhuge2024gptswarm}, \emph{ADAS} \cite{hu2025adas}, and \emph{AgentSquare} \cite{shang2025agentsquare} optimize pipeline graphs or code modules under central control. These systems assume fixed public ground truth and cooperative nodes inside one operator's sandbox, leaving them exposed to silent rule changes, schema-valid hallucinations, and copycat wrappers.

\paragraph{Failure Attribution, Selective Abstention, and Non-Stationary Bandits.}
To debug failed trajectories, \emph{Who\&When} \cite{zhang2025whowhen} feeds full internal agent transcripts to a central LLM judge---an approach blocked by cross-enterprise privacy and high token cost. SYLVA builds instead on team contract theory \cite{holmstrom1982moral}, isolating faults at black-box stage boundaries via output checks, penalty-free abstention (grounded in Chow's reject option \cite{chow1970optimum} and work showing binary grading pushes LLMs to hallucinate \cite{kalai2025hallucinate}), and 3-peer step replay. For online routing, SYLVA adapts Disjoint Linear Upper Confidence Bound (\texttt{LinUCB}) \cite{li2010contextual} contextual bandits with discounted exponential forgetting (\texttt{D-LinUCB} \cite{russac2019weighted}), rookie trial quotas, and fee penalties (Table~\ref{tab:positioning}).

\section{How SYLVA Works \& Why Its Incentives Hold}

\subsection{Step-by-Step Protocol Walkthrough}

\begin{figure}[!t]
\centering
\includegraphics[width=\textwidth]{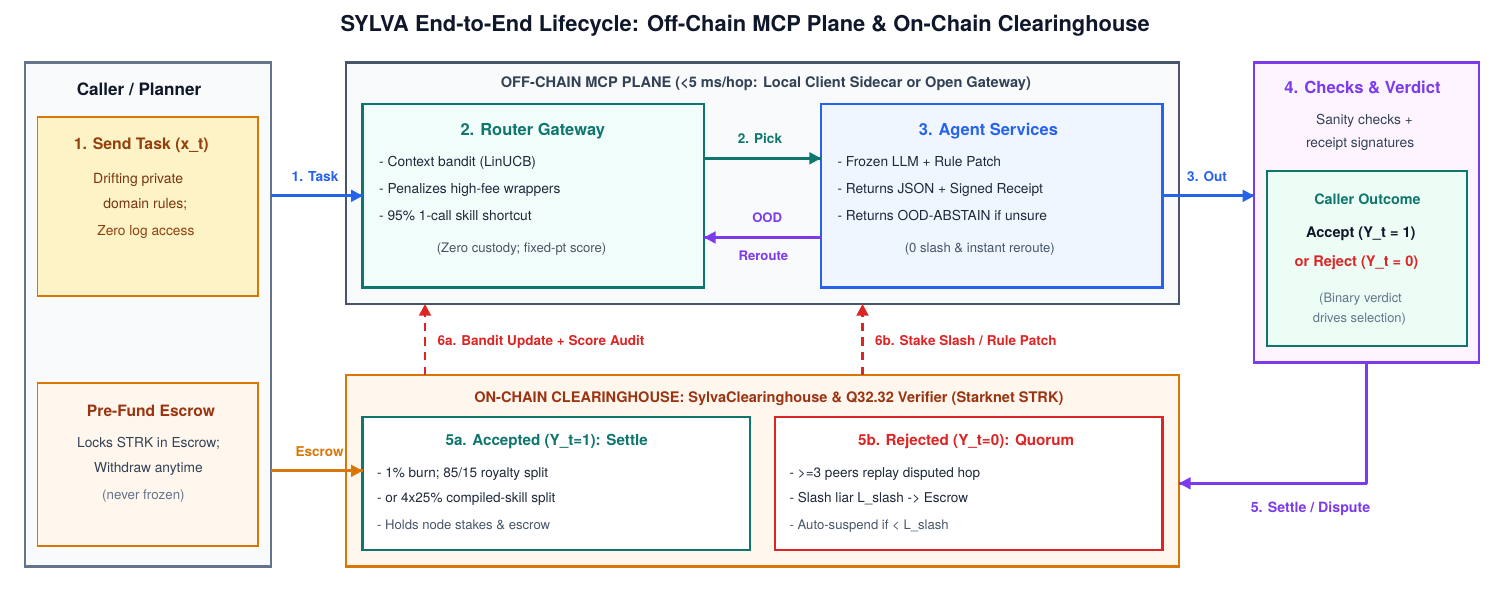}
\caption{SYLVA end-to-end clockwise lifecycle (Steps~1--6): fast off-chain MCP routing, agent services, and boundary checks (run via a caller sidecar or open gateway) pair with binary caller feedback ($Y_t \in \{0,1\}$) and on-chain Starknet (\texttt{STRK} token) settlement, peer-quorum slashing ($L_{\mathrm{slash}}$), and verifiable fixed-point routing audits.}
\label{fig:arch}
\end{figure}

Consider a stream of multi-step tasks $t = 1, \dots, T$ (in our testbed, $H=4$ sequential stages, though the protocol supports any $H \ge 2$). Each agent keeps frozen weights ($\Delta W = 0$), so SYLVA never inspects internal prompts or updates model parameters. Instead, it routes tasks step by step through four components (Figure~\ref{fig:arch}, summarized in Algorithm~1):
\begin{itemize}
    \item \textbf{Router \& Skill Compiler}: Maps each request's discrete category (jurisdiction $k$) and payload into a 12-dimensional normalized fingerprint $x_t$ ($\|x_t\|_2 = 1$). Using two compact per-provider statistics---a precision matrix $A_{t,v} \in \mathbb{R}^{12 \times 12}$ and a reward-weighted feature sum $b_{t,v} \in \mathbb{R}^{12}$, giving estimated quality $\hat{\theta}_{t,v} = A_{t,v}^{-1} b_{t,v}$---the router ranks providers at each stage by predicted quality minus price, plus a bonus for under-tested nodes. When one 4-node team succeeds reliably on category $k$, the router compiles it into a 1-call shortcut ($95\%$ fast path, $5\%$ exploration).
    \item \textbf{Agent Services \& Planner}: Each stage hosts independent providers (regional specialists updated in plain English when rules shift, cheaper generalists, faulty hallucinators, and marked-up wrappers). Task decomposition can be performed either by the caller's own host agent (as in our Starknet deployment) or by competing market planners ranked by the router (as in our Tier-1 and Tier-2 suites, where two reliable planners err at $q=0.02$ and one careless planner errs at $q=0.35$).
    \item \textbf{Boundary Verifier \& Step Replay}: Why can a caller supply $Y_t \in \{0, 1\}$ without knowing private tariff formulas beforehand? Generating a compliant multi-jurisdiction filing requires specialist rules up front, whereas observing whether a downstream clearing rail or audit accepts the finished voucher ($Y_t \in \{0, 1\}$) is directly observable after submission. After each step runs, the verifier checks formatting and numeric sanity. If a specialist declares itself out of domain (\texttt{OOD-ABSTAIN}), the router retries the same stage with zero penalty. When a completed chain is rejected ($Y_t = 0$) despite passing surface sanity checks, the verifier replays disputed steps across $m \ge 3$ independent peers---alongside $5\%$--$10\%$ random shadow probes. Crucially, because each signed receipt commits to the exact intermediate \texttt{input\_hash} received by that step, replaying Stage~2 on a corrupted input from Stage~1 reproduces Stage~2's actual output (exonerating Stage~2), whereas replaying Stage~1 on its clean \texttt{input\_hash} exposes Stage~1 as the culprit.
    \item \textbf{On-Chain Clearinghouse}: Every step produces a signed receipt. Once an accepted episode ($Y_t = 1$) finishes, the clearinghouse pays providers from pre-funded caller escrow, splits compiled-skill fees $4 \times 25\%$, and sends a configurable creator share---set to $85\%$---of registered-wrapper fees back to the original specialist. When an episode is rejected ($Y_t = 0$), innocent steps and the $m \ge 3$ replay peers are still compensated, while the confirmed faulty step's stake slash $L_{\mathrm{slash}}$ ($8\times$--$25\times$ a step fee) credited to caller escrow reimburses both the bad hop and peer-replay costs.
\end{itemize}

\begin{center}
\begin{minipage}{\textwidth}
\small
\begin{tabularx}{\textwidth}{lX}
\toprule
\multicolumn{2}{l}{\textbf{Algorithm 1: Single-Episode SYLVA Loop}} \\
\midrule
\textbf{Input:} & Task category (jurisdiction) $k$, fingerprint $x_t \in \mathbb{R}^{12}$ ($\|x_t\|_2{=}1$), payload $u_t$, stage providers $\{\mathcal{V}_h\}_{h=0}^{H-1}$, router statistics $\{(A_{t,v}, b_{t,v})\}$, clearinghouse $\mathcal{L}$. \\
\textbf{1. Fast Path (Router):} & If category $k$ has an active compiled skill $\tilde{v}_k$ and $\mathrm{rand}() < 0.95$, call $\tilde{v}_k$ once; otherwise run Steps~2--3. \\
\textbf{2. Planner:} & Caller host or chosen market planner issues step instructions (valid with prob.\ $1-q$; faulty spec with prob.\ $q$). \\
\textbf{3. Per-Hop Loop:} & \textbf{for} stage $h = 0, \dots, H-1$ \textbf{do}: \\
 & \quad (a) \textbf{Router} picks provider $v_h \in \mathcal{V}_h$ via bandit score $S_t$ (Eq.~\eqref{eq:ucb}) or $3\%$--$10\%$ rookie (\texttt{canary}) quota $\epsilon_{\text{canary}}$. \\
 & \quad (b) \textbf{Service} runs $v_h$; if it returns \texttt{OOD-ABSTAIN}, $\mathcal{L}$ pays a $5\%$ screening micro-fee ($0.0005$ credits; $0.001\,\mathrm{STRK}$ on-chain) with zero slash and the router retries stage $h$ (up to $R_{\max}{=}4$ candidates / $3$ OOD retries). \\
 & \quad (c) \textbf{Verifier} checks output sanity and runs a random shadow probe with prob.\ $5\%$--$10\%$. \\
\textbf{4. User Check:} & End user / clearing rail accepts ($Y_t = 1$) or rejects ($Y_t = 0$) within tolerance $\tau$. If $Y_t = 0$ despite passing sanity checks, the verifier runs 3-peer replay using each step's signed \texttt{input\_hash}. \\
\textbf{5. Settle \& Learn:} & \textbf{Clearinghouse} settles signed receipts ($85\%/15\%$ royalties or $4 \times 25\%$ skill split) and slashes confirmed faulty nodes ($L_{\text{slash}}$). \textbf{Router} updates $(A_{t,v}, b_{t,v})$ with forgetting $\gamma$ and fee penalty $\lambda_f$, and compiles stable chains into $\tilde{v}_k$. \\
\bottomrule
\end{tabularx}
\end{minipage}
\end{center}
\vspace{0.6em}

\subsection{Routing Decision Rules \& System-2 \texorpdfstring{$\leftrightarrow$}{<->} System-1 Lifecycle}
\label{sec:routing_lifecycle}

Here is how those rules play out on a single incoming invoice---first when the router explores candidates step by step (System~2), and later when a stable team is compiled into a one-call shortcut (System~1 \cite{kahneman2011thinking}).

\paragraph{(A) Four Routing Decision Rules in System-2 Exploration.}
Feature hashing turns \texttt{(domain, input)} into the 12-number fingerprint $x_t$. For every provider $v$, the router initializes a ridge-regularized uncertainty matrix $A_{0,v} = \lambda I_{12}$ ($\lambda = 1.0$) and reward sum $b_{0,v} = 0$ (starting from an unpulled expected score of $0.0$). At each stage $h \in \{0,\dots,H-1\}$, it picks among active providers $v \in \mathcal{V}_h$ via four rules:
\begin{enumerate}
    \item \textbf{Rookie (\texttt{canary}) Trial Quota ($\epsilon_{\mathrm{canary}}$):} If candidates with fewer than $N_{\mathrm{canary}}$ calls exist ($4$--$10$), the router picks one at random $3\%$--$10\%$ of the time so newly patched specialists always get a fair trial.
    \item \textbf{Score = Expected Net Quality + Curiosity Bonus:} Otherwise, each candidate is scored by
    \begin{equation}
    \label{eq:ucb}
        S_t(v, x_t) = \underbrace{\hat{\theta}_{t,v}^\top x_t}_{\text{predicted net quality}} + \underbrace{\kappa \sqrt{x_t^\top A_{t,v}^{-1} x_t}}_{\text{unfamiliarity bonus}},
    \end{equation}
    where $\hat{\theta}_{t,v} = A_{t,v}^{-1} b_{t,v}$ predicts node $v$'s net performance on task type $x_t$, and the second term shrinks as $v$ gains experience on similar inputs while staying large on unfamiliar tasks.
    \item \textbf{Instant Reroute on \texttt{OOD-ABSTAIN}:} If a node says it cannot handle this input, the clearinghouse charges zero slash and pays a $5\%$ screening micro-fee. In simulation (Tiers~1--2), the router records a $+0.08$ credit on each clean abstention; on Starknet (Tier-3), it applies a neutral $0.0$ post-warmup router reward and enforces a hard $40\%$ abstention cap after 3 warmup calls so dummy nodes cannot farm screening fees. Either way, the router immediately excludes that node and picks the next best candidate on the \emph{same} stage.
    \item \textbf{Fee Penalty, Step-Only Blame, and Forgetting ($\gamma$):} On each update $r$, older statistics decay by $\gamma \in (0, 1]$ via $b_{t+1,v} = \gamma b_{t,v} + r x_t$ and $A_{t+1,v} = \gamma A_{t,v} + x_t x_t^\top + (1-\gamma)\lambda I_{12}$, flushing stale incumbents after rule shifts. Bandit targets $r$ preserve a strict incentive order around the $0.0$ untested prior: accepted run ($+1.0 - \lambda_f f_v$, pricing out marked-up wrappers) $>$ innocent teammate on a disputed run ($+0.25$) $>$ honest OOD abstention ($+0.08$ / $0.0$) $\ge$ untested node ($0.0$) $>$ unresolved stealth failure ($-0.12$, zero slash) $>$ isolated faulty worker or planner ($-0.60$ / $-0.65$).
\end{enumerate}

\paragraph{(B) How 4-Hop System-2 Compiles into 1-Call System-1---and Reverts on Drift.}
Running four separate LLM calls per task wastes tokens because every stage re-reads the same raw invoice ($2,765.8$ tokens/task in Tier-2). SYLVA moves between 4-step exploration (System~2) and 1-call shortcuts (System~1) in four steps:
\begin{enumerate}
    \item \textbf{Compilation Trigger:} \texttt{SkillConsolidator} counts how often the exact same quartet $[v_0, v_1, v_2, v_3]$ succeeds in category $k$. Once it hits $\ge N_{\min}$ runs at $\ge p_{\min}$ pass rate ($18$ runs / $90\%$ in Tier-1; $8$ runs / $88\%$ in Tier-2), the router creates a compiled skill $\tilde{v}_k$.
    \item \textbf{Physical Execution (Why 4 Calls Become 1):} A compiled skill collapses four round-trips into one call in either of two ways: (i)~\emph{Merged Single-Prompt Skill} (Tier-2 \texttt{LLMSkillConsolidator}), combining the four specialists' plain-text rules into one prompt specialized solely for category $k$ so the LLM reads the invoice only once ($600.8$ tokens/task, $-78.3\%$, abstaining outside category $k$), applicable when one operator, a consenting consortium, or a trusted execution enclave (TEE) merges rulebooks; or (ii)~\emph{Pre-Wired 4-Node Pipeline} (Tier-3 cross-company commercial tools), where one lead parent coordinates the quartet behind a single endpoint, preserves full prompt secrecy, and signs one \texttt{EXECUTED\_SKILL} receipt.
    \item \textbf{4$\times$25\% On-Chain Revenue Split:} Registering $\tilde{v}_k$ on Starknet locks all four parent addresses $[v_0, v_1, v_2, v_3]$ into the \texttt{skills} table and automatically splits $100\%$ of net skill fees equally ($25\%$ each), so specialists earn steady dividends rather than losing revenue to shortcuts.
    \item \textbf{Drift Circuit Breaker \& Fallback to System-2:} The router sends $95\%$ of in-context traffic to $\tilde{v}_k$ while keeping $5\%$ on regular 4-step exploration. Meanwhile, \texttt{SylvaClearinghouse} watches a sliding win-rate window on $\tilde{v}_k$: if external rule drift drops accuracy below $80\%$ over $\ge 5$ calls, the contract sets \texttt{active = false}, immediately returning all traffic to 4-step System-2 exploration until patched specialists form a new winning team.
\end{enumerate}

\subsection{Formal Incentive Bounds}

Table~\ref{tab:notation} summarizes the protocol and incentive-analysis notation used below.

\begin{table}[!t]
\centering
\caption{Summary of protocol and incentive-analysis notation.}
\label{tab:notation}
{\footnotesize\setlength{\tabcolsep}{4pt}
\begin{tabularx}{\textwidth}{l>{\raggedright\arraybackslash}Xc}
\toprule
Symbol & Meaning & Simulation / On-Chain Value \\
\midrule
$x_t,\, k,\, d$ & Normalized task fingerprint; discrete category; dimension & $\|x_t\|_2{=}1$;\, $k \in \{0..7\}$;\, $d{=}12$ \\
$Y_t,\, \tau$ & Binary caller verdict; end-to-end numeric tolerance & $Y_t \in \{0,1\}$;\, $10^{-3}$ (T1) / $\$0.05$ (T2) \\
$p,\, r_v,\, L$ & Subjective pass prob.; passing bandit utility; failure loss & $p \in [0,1]$;\, $r_v{=}1.0$;\, $L{=}0.60$ \\
$L_{\mathrm{slash}},\, u_{\mathrm{abs}}$ & On-chain stake slash; abstention bandit utility & $0.08$ / $0.50\,\mathrm{STRK}$;\, $+0.08$ / $0.0$ \\
$q,\, \epsilon,\, F$ & Planner error prob.; combined sibling fault prob.; $q{+}(1{-}q)\epsilon$ & $q \in \{0.02, 0.35\}$;\, $\epsilon, F \in (0,1)$ \\
$\alpha_{\mathrm{fp}},\, \beta_{\mathrm{fn}},\, \beta_{\mathrm{tox}}$ & Sanity false-alarm rate; normal miss rate; stealth evasion rate & $0.02$;\, $0.12$;\, $0.60$ \\
$\eta_{\mathrm{miss}}$ & Residual unresolved stealth-fault fraction after replay & $\eta_{\mathrm{miss}} \in [0,1)$ \\
$r_{\mathrm{prot}},\, r_{\mathrm{pen}},\, r_{\mathrm{unres}}$ & Protected sibling credit; isolated fault penalty; unresolved damping & $+0.25$;\, $0.60$;\, $0.12$ (zero slash) \\
$f_v,\, \Delta f,\, \mu_v(x)$ & Node fee; wrapper markup ($f_w{-}f_v$); expected quality $\mathbb{E}[r_{t,v}|x]$ & $0.010$ / $0.02\,\mathrm{STRK}$;\, $\Delta f > 0$;\, $[0,1]$ \\
$\lambda_f,\, \rho_{\mathrm{core}}$ & Fee-penalty weight; configurable wrapper creator royalty share & $18.0$ (wrapper flood);\; $\rho_{\mathrm{core}}{=}0.85$ \\
$\kappa,\, \gamma,\, \lambda,\, \epsilon_{\text{canary}}$ & UCB weight; forgetting factor; ridge init.; rookie quota & $0.40$--$0.45$;\, $1.0/0.988/0.96$;\, $1.0$;\, $3$--$10\%$ \\
\bottomrule
\end{tabularx}}
\end{table}

\paragraph{1. Keeping Specialists Honest via Penalty-Free Abstention.}

\noindent\textbf{Takeaway.} As shown in selective classification \cite{chow1970optimum} and recent analyses of binary LLM grading \cite{kalai2025hallucinate}, punishing ``I don't know'' as harshly as a wrong answer forces rational agents to guess blindly. Giving safe OOD abstention a small cushion creates an interior confidence threshold below which honest agents voluntarily step aside.

\begin{proposition}[Incentive Compatibility of OOD Abstention]
\label{prop:abstain}
Let a node $v$ face a task where its subjective probability of passing verification is $p \in [0, 1]$. Suppose passing yields normalized utility $r_v > 0$, failing incurs failure loss $L > 0$, and exercising safe OOD abstention yields utility $u_{\mathrm{abs}}$. Then:
\begin{enumerate}
    \item[(a)] Attempting execution strictly dominates abstention if and only if
    \begin{equation}
    \label{eq:abstain_cutoff}
        p > p^\star := \frac{L + u_{\mathrm{abs}}}{r_v + L}.
    \end{equation}
    \item[(b)] Whenever $-L < u_{\mathrm{abs}} < r_v$, there is an interior abstention cutoff $p^\star \in (0, 1)$, so honest specialists abstain cleanly on low-confidence out-of-domain inputs ($p < p^\star$).
    \item[(c)] Conversely, punishing abstention as hard failure ($u_{\mathrm{abs}} \le -L$, tested as an ablation in Section~5) collapses $p^\star \le 0$, forcing rational nodes to guess blindly across all OOD inputs.
\end{enumerate}
\end{proposition}

\begin{proof}
Attempted execution has expected utility $\mathbb{E}[U(\mathrm{attempt})] = p r_v - (1-p)L = p(r_v + L) - L$. Comparing this strictly against abstention utility $u_{\mathrm{abs}}$ and dividing by $r_v + L > 0$ gives $p > \frac{L + u_{\mathrm{abs}}}{r_v + L} = p^\star$, proving~(a). Parts~(b) and~(c) follow immediately: $0 < L + u_{\mathrm{abs}} < r_v + L$ iff $-L < u_{\mathrm{abs}} < r_v$, whereas $L + u_{\mathrm{abs}} \le 0$ makes $\mathbb{E}[U(\mathrm{attempt})] > u_{\mathrm{abs}}$ for every $p > 0$.
\end{proof}
\noindent\emph{Instantiation \& Anti-Abstention-Farming Note.} In normalized bandit utility units ($r_v = +1.0$, $L = 0.60$, $u_{\mathrm{abs}} = +0.08$), Eq.~\eqref{eq:abstain_cutoff} sets the abstention cutoff at $p^\star = (0.60 + 0.08)/(1.0 + 0.60) = 42.5\%$; in raw settlement units, abstaining avoids stake slashing ($0$ vs.\ $L_{\mathrm{slash}}$) while receiving a $5\%$ screening micro-fee ($0.0005$ credits / $0.001\,\mathrm{STRK}$). Because uncapped positive abstention rewards would invite dummy nodes to abstain on every call, our Tier-3 clearinghouse enforces a $40\%$ post-warmup abstention cap and neutral ($0$) post-warmup router reward, blocking abstention farming while preserving an interior cutoff $p^\star = \frac{L}{r_v + L} = 37.5\% > 0$.

\paragraph{2. Preventing Blame Collapse Across Multi-Step Chains.}

\noindent\textbf{Takeaway.} Consistent with Holmstr{\"o}m's team moral hazard \cite{holmstrom1982moral}, when every hop in a failed pipeline is punished equally, even a 100\%-accurate specialist averages a negative score once other stages fail more than half the time ($F > 50\%$). Step-level fault isolation (\texttt{bisect} in code) isolates errors so compliant specialists stay strictly positive regardless of $F$.

\begin{proposition}[Uniform Blame Collapse vs.\ Step-Level Fault Isolation]
\label{prop:bisect}
Consider an $H$-stage chain containing a zero-error specialist $v^\star$ at stage $h$. Let $F := q + (1-q)\epsilon \in (0, 1)$ denote the total probability that another stage fails---either via an upstream specification error (prob.~$q$) or via at least one sibling stage failing given a valid specification (combined conditional prob.~$\epsilon$). Under symmetric whole-chain blame ($+1$ if all stages succeed, $-1$ if any fails):
\begin{equation}
    \mathbb{E}\big[R_{\mathrm{uniform}}(v^\star)\big] = 1 - 2F,
\end{equation}
which turns strictly negative whenever $F > \tfrac{1}{2}$, driving accurate specialists out of selection. Under SYLVA's Step-Level Fault Isolation---with local false-alarm rate $\alpha_{\mathrm{fp}}$, residual unresolved replay-miss rate $\eta_{\mathrm{miss}} \in [0, 1)$ (mild zero-slash damping $-r_{\mathrm{unres}}$), sibling protection reward $r_{\mathrm{prot}} > 0$, and fault penalty $r_{\mathrm{pen}} > 0$---$v^\star$ guarantees uniformly over all $F \in (0, 1)$:
\begin{equation}
    \mathbb{E}\big[R_{\mathrm{isolate}}(v^\star)\big] \ge (1-\alpha_{\mathrm{fp}})\big[(1-\eta_{\mathrm{miss}})r_{\mathrm{prot}} - \eta_{\mathrm{miss}} r_{\mathrm{unres}}\big] - \alpha_{\mathrm{fp}} r_{\mathrm{pen}}.
\end{equation}
In particular, $\mathbb{E}[R_{\mathrm{isolate}}(v^\star)] > 0$ whenever $\alpha_{\mathrm{fp}} < \frac{r_{\mathrm{prot}}}{r_{\mathrm{prot}} + r_{\mathrm{pen}}}$ in the resolved-replay regime ($\eta_{\mathrm{miss}} = 0$), and more generally when $\eta_{\mathrm{miss}} < \frac{(1-\alpha_{\mathrm{fp}})r_{\mathrm{prot}} - \alpha_{\mathrm{fp}} r_{\mathrm{pen}}}{(1-\alpha_{\mathrm{fp}})(r_{\mathrm{prot}} + r_{\mathrm{unres}})}$.
\end{proposition}

\begin{proof}
Since stage $h$ succeeds locally with probability~$1$, whole-chain success occurs with probability $1 - F$ and failure with probability $F$, so $\mathbb{E}[R_{\mathrm{uniform}}(v^\star)] = (+1)(1-F) + (-1)F = 1 - 2F < 0$ iff $F > \tfrac{1}{2}$. Under SYLVA's step-level fault isolation, two mutually exclusive events determine $v^\star$'s reward independently of how often external stages fail ($F$). (1)~With probability $\alpha_{\mathrm{fp}}$, a false alarm on $v^\star$'s local sanity check yields at worst penalty $-r_{\mathrm{pen}}$. (2)~With probability $1 - \alpha_{\mathrm{fp}}$, $v^\star$ passes its local sanity check and earns $+1 \ge r_{\mathrm{prot}}$ if the chain succeeds, $+r_{\mathrm{prot}}$ if peer replay isolates another stage's fault, or $-r_{\mathrm{unres}}$ on the at-most $\eta_{\mathrm{miss}}$ fraction of runs where a stealth sibling fault remains unresolved. Combining both branches gives expected reward at least $(1-\alpha_{\mathrm{fp}})\big[(1-\eta_{\mathrm{miss}})r_{\mathrm{prot}} - \eta_{\mathrm{miss}} r_{\mathrm{unres}}\big] - \alpha_{\mathrm{fp}} r_{\mathrm{pen}}$, which is strictly positive iff $(1-\alpha_{\mathrm{fp}})\big[(1-\eta_{\mathrm{miss}})r_{\mathrm{prot}} - \eta_{\mathrm{miss}} r_{\mathrm{unres}}\big] > \alpha_{\mathrm{fp}} r_{\mathrm{pen}}$, yielding the stated $\eta_{\mathrm{miss}}$ threshold. With values $(r_{\mathrm{prot}}, r_{\mathrm{pen}}, r_{\mathrm{unres}}, \alpha_{\mathrm{fp}}) = (0.25, 0.60, 0.12, 0.02)$, positive expected credit holds for all $F \in (0,1)$ whenever replay resolves at least $35.7\%$ of stealth sibling disputes ($\eta_{\mathrm{miss}} < 64.3\%$).
\end{proof}

\paragraph{3. Driving Out Wrappers \& Protecting Creator Revenue.}

\noindent\textbf{Takeaway.} Thin wrappers match a core specialist's accuracy while charging a higher price. Subtracting scaled fees directly from LinUCB rewards ensures wrappers lose routing score once uncertainty shrinks below the fee gap, while clearinghouse royalties return a configurable creator share (set to $85\%$ in our experiments) on any stray calls.

\begin{proposition}[Wrapper Elimination and Core Revenue Retention]
\label{prop:wrapper}
Suppose the Disjoint LinUCB router trains on fee-penalized net rewards $\tilde{r}_{t,i} = r_{t,i} - \lambda_f f_i$ ($\lambda_f > 0$) and scores arms via Eq.~\eqref{eq:ucb}, while the clearinghouse enforces a configurable creator royalty share $\rho_{\mathrm{core}} \in (0, 1]$ (set to $85\%$ in our experiments). Consider an in-protocol wrapper $w$ copying core specialist $v$ with no higher expected pre-fee quality ($\mu_w(x) := \mathbb{E}[r_{t,w}\mid x] \le \mu_v(x)$) and strict markup $\Delta f := f_w - f_v > 0$. Then:
\begin{enumerate}
    \item[(a)] In the stationary regime ($\gamma = 1$) across any recurrent discrete category $x$ (such as our $K=8$ jurisdictions), $\limsup_{t\to\infty} \big[S_t(w, x) - S_t(v, x)\big] \le -\lambda_f \Delta f < 0$ (i.e., once exploration uncertainty shrinks, wrapper $w$'s score stays permanently below specialist $v$'s score by at least the scaled markup $\lambda_f \Delta f$), bounding non-rookie UCB pulls of $w$ at $O\big((\lambda_f \Delta f)^{-2} \log T\big)$.
    \item[(b)] On every lineage-registered call routed to $w$, the clearinghouse atomically transfers fraction $\rho_{\mathrm{core}}$ ($85\%$ of net post-burn fee on Starknet, or $84.15\%$ of gross spend $f_w$, exceeding $100\%$ of $f_v$ whenever $f_w > 1.19 f_v$) to creator $v$, preserving $\ge \rho_{\mathrm{core}}$ core revenue regardless of exploration or forgetting transients.
\end{enumerate}
\end{proposition}

\begin{proof}
Under standard Disjoint LinUCB ($\gamma = 1$) on unit-norm context vectors \cite{li2010contextual}, both the mean-prediction error and the exploration bonus $\kappa \sqrt{x^\top A_{t,i}^{-1} x}$ shrink at rate $O(\sqrt{(\log t)/N})$ after $N$ pulls in context $x$. Because the expected net reward gap satisfies $\mathbb{E}[\tilde{r}_{t,w} - \tilde{r}_{t,v}] = (\mu_w(x) - \mu_v(x)) - \lambda_f \Delta f \le -\lambda_f \Delta f < 0$, once both arms reach $N \gtrsim O\big(\kappa^2 (\lambda_f \Delta f)^{-2} \log T\big)$ context observations, the fixed fee gap $-\lambda_f \Delta f$ dominates the confidence radii, guaranteeing $S_t(w,x) < S_t(v,x)$ asymptotically and proving~(a). Part~(b) follows by contract construction: \texttt{settle\_episode} routes $\rho_{\mathrm{core}}$ of net fee $f_w$ directly to the creator balance whenever parent $v$ is registered.
\end{proof}
\begin{remark}[Stationary vs.\ Non-Stationary Forgetting Regime]
Part~(a) analyzes $\gamma = 1$ (used in our wrapper-flood experiment, Section~5). When rule drift requires discounted forgetting ($\gamma < 1$, D-LinUCB \cite{russac2019weighted}), effective sample memory caps at $O((1-\gamma)^{-1})$ and periodically re-probes unselected arms; the clearinghouse's unconditional $85\%$ creator split in~(b) protects original creators continuously throughout those re-exploration transients.
\end{remark}

\subsection{Five Practical Attacks and How SYLVA Stops Them}
\label{sec:threat_model}

Open agent markets invite strategic cheating across workers, middlemen, and routers. SYLVA blocks five concrete attacks:
\begin{enumerate}
    \item \textbf{Confident Schema-Valid Hallucinators.} \emph{Attack:} An agent returns valid JSON filled with fabricated numbers, slipping past surface sanity checks $60\%$ of the time ($\beta_{\mathrm{tox}} = 0.60$). \emph{Defense:} SYLVA runs $5\%$--$10\%$ random shadow probes alongside 3-peer replay whenever a caller rejects a run, slashing faulty nodes' on-chain stake. Once remaining stake drops below $L_{\mathrm{slash}}$, the Starknet contract automatically deactivates the node (\texttt{active = false}) so attackers cannot keep running with zero remaining stake at risk.
    \item \textbf{Marked-Up Wrappers (Registered or Hidden).} \emph{Attack:} A middleman resells an existing specialist's work at a higher price $f_w = f_v + \Delta f$, either registering its parent link on-chain or hiding it. \emph{Defense:} Registered wrappers are starved by router fee penalties ($\lambda_f > 0$, Proposition~\ref{prop:wrapper}) while still paying $85\%$ creator royalties on any stray call. Unlinked high-price proxies are likewise priced out by $\lambda_f$; stopping off-chain wholesale arbitrage completely requires core providers to require signed lineage receipts for bulk API access.
    \item \textbf{Sybil Wash-Trading Rings.} \emph{Attack:} Colluding accounts call one another repeatedly and post five-star ratings to hijack leaderboards. \emph{Defense:} Star ratings never enter routing decisions, which learn solely from settled caller outcomes $(x_t, Y_t)$. In addition, the clearinghouse burns a fixed settlement fee per handoff ($1\%$, or \texttt{100} basis points [\texttt{bps}] where $1\,\text{bps} = 0.01\%$, sent to an unwithdrawable burn sink on Starknet), making circular wash trades strictly unprofitable.
    \item \textbf{Malicious Planner Extortion.} \emph{Attack:} A compromised planner intentionally sends corrupted instructions so downstream specialists fail sanity checks and lose stake. \emph{Defense:} Slashed stakes go strictly to refund the affected caller's escrow---never to the planner---and 3-peer replay shifts blame back to the planner ($-0.65$ router penalty).
    \item \textbf{Biased or Bribed Router Gateways.} \emph{Attack:} A compromised router gateway secretly boosts scores for kickback-paying wrappers or starves competing specialists without touching on-chain settlement. \emph{Defense:} The off-chain Rust router and Cairo \texttt{verifiable\_router} contract share bit-exact deterministic 32.32-bit fixed-point (\texttt{Q32.32}) arithmetic (\texttt{score\_arm\_q32}) \cite{bensasson2018stark,goldberg2021cairo}, letting any caller verify bandit scores and $\arg\max$ choices on-chain without exposing proprietary node prompts.
\end{enumerate}

\section{Benchmark Setup: Four-Stage Invoice Settlement}
\label{sec:poc_scope}

We evaluate SYLVA on a reproducible four-stage cross-domain invoice settlement benchmark across $K=8$ regulated jurisdictions (such as \texttt{EU-MedDevice}, \texttt{US-SaaS-Tax}, \texttt{UK-FinClear}, and \texttt{CA-Pharma}). Each task provides a $d=12$ context vector---combining an $8$-slot jurisdiction indicator with $4$ scaled numeric features (order magnitude, foreign-exchange [FX] rate, strictness, and urgency), normalized to $\|x_t\|_2 = 1$---and requires four dependent calculations:
\begin{enumerate}
    \item \textbf{Stage 0 (FX \& Schema Cleaning)}: $a_{\mathrm{clean}} = \mathrm{raw\_amount} \times \mathrm{fx\_rate}$.
    \item \textbf{Stage 1 (Tariff \& Compliance)}: $t_{\mathrm{fee}} = a_{\mathrm{clean}} \cdot r^{\mathrm{tariff}}_k \cdot \sigma_t + b_k$, where $\sigma_t \in \mathbb{R}$ captures upstream instruction quality ($\sigma_t = 1$ when valid; $\sigma_t \ne 1$ when the upstream planner gives bad input).
    \item \textbf{Stage 2 (Inventory Risk Reserve)}: $\rho_{\mathrm{res}} = (a_{\mathrm{clean}} + t_{\mathrm{fee}}) \cdot r^{\mathrm{risk}}_k$.
    \item \textbf{Stage 3 (Net Settlement Voucher)}: $v_{\mathrm{net}} = a_{\mathrm{clean}} + t_{\mathrm{fee}} + \rho_{\mathrm{res}}$.
\end{enumerate}
Why fix $H=4$ in our testbed (whereas the Tier-3 gateway supports arbitrary stage counts)? These four stages capture a realistic invoice-settlement workflow while providing the minimal sequential depth needed to exercise every mechanism simultaneously: compounding $72\%$-accurate generalist errors ($0.72^4 \approx 26.9\%$), multi-hop blame spillover (one faulty step dragging down three innocent teammates under uniform scoring), and $4 \to 1$ skill compilation with $4 \times 25\%$ payouts. In Tiers~1--2, honest providers charge a uniform base step fee of $f_v = 0.010$ credits and post $1.0$ credit of initial stake (stealth hallucinators undercut at $0.008$); an \texttt{OOD-ABSTAIN} pays a $0.0005$-credit ($5\%$) screening micro-fee with zero slash, whereas isolated execution faults incur a $0.08$-credit ($8\times$ step fee) stake slash. Outside wrapper-flood tests (where copycat markups introduce price dispersion), equal base fees make $\lambda_f = 0.0$ equivalent to $\lambda_f > 0$. The local statutory rates $(r^{\mathrm{tariff}}_k, b_k, r^{\mathrm{risk}}_k)$ represent private regional rules missing from static model pretraining. To simulate silent regulatory shifts (in our rule-drift experiments in Sections~5 and~6), half of the eight jurisdictions undergo unannounced mutations ($\Delta r^{\mathrm{tariff}} = +0.055$, $\Delta b = +14.0$, $\Delta r^{\mathrm{risk}} = +0.022$).

\paragraph{What Is Simulated in Tiers~1--2 vs.\ Live On-Chain in Tier~3.}
For transparency, our multi-seed Python sweeps (Tiers~1--2) use three fast simulation helpers to run hundreds of episodes cheaply, whereas our Rust/Cairo production implementation (Tier-3, Section~\ref{sec:tier3_starknet}) enforces zero-oracle checks live on Starknet:
\begin{itemize}
    \item \emph{Simulated Stage Checks vs.\ On-Chain Peer Quorum}: In Tiers~1--2, the POC script (\texttt{poc/protocol.py}) uses identical noisy check parameters ($\beta_{\mathrm{fn}} = 0.12, \beta_{\mathrm{tox}} = 0.60, \alpha_{\mathrm{fp}} = 0.02$) and sandbox stage expectations when running replay or tagging $\sigma_t \ne 1$, keeping the comparison between rule stubs (Tier-1) and live LLMs (Tier-2) strictly apples-to-apples. Section~\ref{sec:tier3_starknet} removes these simulation hooks completely via $m \ge 3$ peer-quorum consensus on Starknet, and Section~\ref{sec:human_study} marks full human/quorum benchmark sweeps as future work.
    \item \emph{Skill Compilation \& Equal Revenue Split}: When the router compiles a stable 4-hop chain into a single-call skill, it merges only the participating specialists' prompt rules (without reading hidden ground truth) and splits net fees evenly ($25\%$ each among the four parents, vs.\ $85\%/15\%$ for single-node wrappers), targeting consenting groups or trusted execution enclaves (TEEs).
    \item \emph{Full Crypto vs.\ Fast Ledger Mode}: Unit tests and the 24-episode interactive CLI console (\texttt{poc/interactive\_symbiont.py --demo}) compute real Ed25519 signatures and flush every write to SQLite. Multi-seed benchmark sweeps use \texttt{fast\_crypto=True, buffered=True} (keyed BLAKE2b HMAC receipts and batched SQLite commits) to keep large sweeps fast.
\end{itemize}
For Tier-2 frozen-LLM experiments, tasks are rendered as noisy commercial memos with distractor credit limits and 30-day payment-term (\texttt{Net-30}) penalty clauses, evaluated within $\tau = \$0.05$ to accommodate cent-level rounding across four LLM steps (vs.\ exact formula tolerance $\tau = 10^{-3}$ in Tier-1). Protocol hyperparameters across both tiers appear in Table~\ref{tab:hyperparams}.

\section{Experiments I: Large-Population Mechanism Dynamics (Tier-1)}

\begin{figure}[!htbp]
\centering
\includegraphics[width=\textwidth]{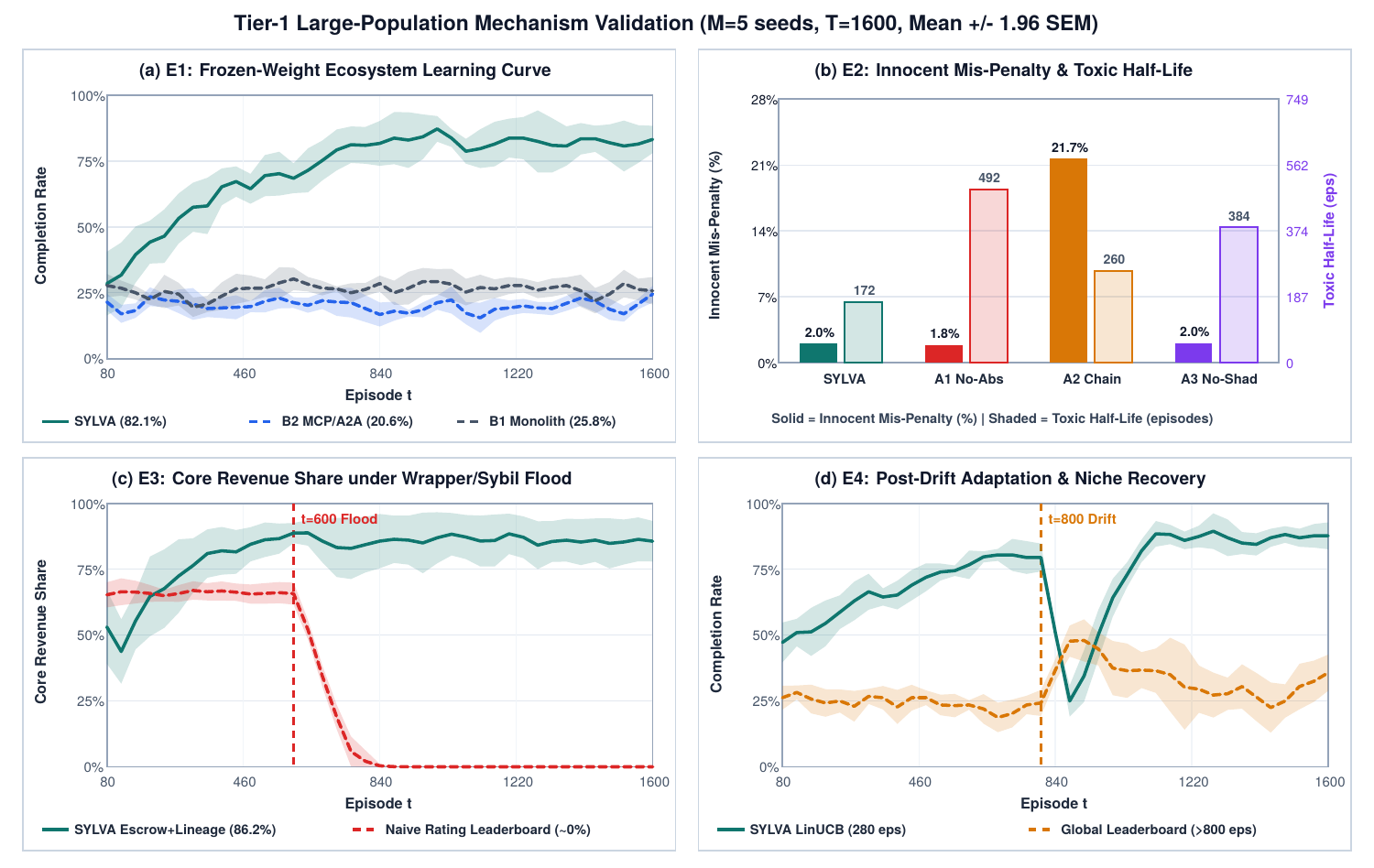}
\caption{Tier-1 mechanism dynamics across $M=5$ seeds ($T=1600$ episodes; line curves show $80$-episode rolling averages starting at $t=80$, and shaded bands show $\mathrm{Mean} \pm 1.96\,\mathrm{SEM}$): \textbf{(a) E1 Learning Curve}---end-to-end completion vs.\ episode $t$, showing SYLVA climbing to $82.1\%$ with frozen weights while monolith ($\mathrm{B1}$, $25.8\%$) and unattributed MCP/A2A ($\mathrm{B2}$, $20.6\%$) stay flat; \textbf{(b) E2 Attribution Ablation}---dual-axis comparison of false penalties on innocent specialists (left axis, solid bars; lower is better) and the earliest $60$-episode window suppressing toxic hallucinator traffic to $\le 5\%$ (right axis, shaded bars, labeled \texttt{Toxic Half-Life}; lower is better), showing uniform blame ($\mathrm{A2}$) spikes false penalties $10.8\times$ ($21.7\%$) while removing abstention ($\mathrm{A1}$) or shadow probes ($\mathrm{A3}$, labeled \texttt{No-Shad}) stalls toxic clearance ($492$ and $384$ eps); \textbf{(c) E3 Wrapper/Sybil Flood}---core creator share of total market revenue before and after injecting $120$ wrappers $+$ $12$ Sybils at $t=600$ (vertical dashed line), where rating leaderboards collapse to $\approx 0\%$ while SYLVA retains $86.2\%$; \textbf{(d) E4 Silent Rule Drift}---completion rate across an unannounced statutory shift in half of all jurisdictions at $t=800$ (vertical dashed line), where SYLVA recovers to $88.0\%$ within $280$ episodes via forgetting ($\gamma=0.988$) while a non-forgetting global leaderboard remains stuck at $30.6\%$.}
\label{fig:tier1}
\end{figure}

Tier-1 tests market dynamics in a 64-agent synthetic ecosystem ($16$ agents per stage: $8$ regional specialists, $4$ generalists at $\approx 0.72$ single-hop accuracy, and $4$ stealth hallucinators with $\beta_{\mathrm{tox}} = 0.60$), paired with two reliable planners ($q=0.02$) and one careless planner ($q=0.35$). Combined with $98.5\%$ synthetic per-hop specialist accuracy ($0.985^4 \approx 94.1\%$), ongoing exploration across $16$ candidates per stage, and $q=0.02$ benign-planner errors, full-chain completion in Tier-1 converges near $82\%$--$83\%$ (whereas deterministic temperature-$0.0$ LLM specialists in Tier-2's compact $8$-node stages reach the $\approx 98\%$ benign-planner ceiling). We evaluate five suites over $M=5$ random seeds (\texttt{[42, 101, 202, 303, 404]}) for $T=1600$ episodes at $\tau = 10^{-3}$ (wall-clock runtime $37.42\,\mathrm{s}$ in fast-ledger simulation mode): \textbf{$\mathrm{E1}$}~(ecosystem learning curve), \textbf{$\mathrm{E2}$}~(fault-isolation ablations), \textbf{$\mathrm{E3}$}~(wrapper and Sybil flood, expanding to $196$ agents), \textbf{$\mathrm{E4}$}~(silent rule drift, expanding to $80$ agents), and \textbf{$\mathrm{E5}$}~(1-call skill compilation). Reported intervals give $\mathrm{Mean} \pm 1.96$ standard error of the mean ($\mathrm{SEM}$) over the converged tail $t \in [1400, 1600)$. Full trajectories appear in Figure~\ref{fig:tier1}.

\paragraph{Panel (a): E1---Can frozen specialists beat a monolith?}
Figure~\ref{fig:tier1}(a) and Table~\ref{tab:tier1_e1} compare end-to-end completion across $t=80..1600$ against a standalone monolithic generalist ($\mathrm{B1}$, flat near the $0.72^4 \approx 26.9\%$ compounding ceiling) and unattributed static MCP/A2A matching ($\mathrm{B2}$). While $\mathrm{B1}$ and $\mathrm{B2}$ stay flat at $25.8\%$ and $20.6\%$, SYLVA learns context-specific chains online and climbs to $\mathbf{82.1\% \pm 6.3\%}$ without changing model weights.

\paragraph{Panel (b): E2---Why safe abstention, hop fault isolation, and shadow probes are all needed.}
Figure~\ref{fig:tier1}(b) uses dual axes to track two failure modes simultaneously (lower is better for both): solid bars (left axis) report how often accurate specialists get falsely penalized, while shaded bars (right axis, labeled \texttt{Toxic Half-Life}) report the earliest $60$-episode window where rolling toxic hallucinator traffic drops from its $25\%$ pool share to $\le 5\%$ (Table~\ref{tab:tier1_e2}). Removing any defense breaks one side:
\begin{itemize}
    \item \emph{No Safe Abstention ($\mathrm{A1}$)}: Punishing abstention like failure forces agents to guess OOD, stretching toxic clearance by $\mathbf{2.86\times}$ ($492.0$ vs.\ $172.0$ episodes).
    \item \emph{No Random Shadow Probes ($\mathrm{A3}$)}: Disabling probes lets stealth hallucinators hide inside valid schemas, more than doubling toxic persistence ($384.0$ episodes, $\mathbf{2.23\times}$).
    \item \emph{Uniform Chain Blame ($\mathrm{A2}$)}: Blaming all four hops equally spikes false penalties on innocent specialists by $\mathbf{10.8\times}$ ($21.67\%$ vs.\ $2.00\%$), dragging specialist scores below unpulled priors and collapsing completion to $13.7\% \pm 7.7\%$. Only full \textbf{SYLVA} keeps both metrics low ($2.0\%$ mis-penalty, $172$ episodes to suppress toxic traffic).
\end{itemize}

\begin{table}[!t]
\centering
\caption{Tier-1 $\mathrm{E1}$ ($M=5, T=1600$): End-to-end completion with frozen model weights ($\Delta W = 0$, $\mathrm{Mean} \pm 1.96\,\mathrm{SEM}$).}
\label{tab:tier1_e1}
{\small\setlength{\tabcolsep}{7pt}
\begin{tabular}{lcc}
\toprule
Method & Tail Completion Rate (\%, no manual edits) & 1.96-SEM Band \\
\midrule
\textbf{SYLVA Complete} & $\mathbf{82.1 \pm 6.3}$ & $[75.8, 88.4]$ \\
B1 Monolithic Generalist ($\approx 0.72^4$) & $25.8 \pm 1.8$ & $[24.0, 27.6]$ \\
B2 Static MCP/A2A Matching \cite{anthropic2024mcp,google2025a2a} & $20.6 \pm 1.4$ & $[19.2, 22.0]$ \\
\bottomrule
\end{tabular}}
\end{table}

\begin{table}[!t]
\centering
\caption{Tier-1 $\mathrm{E2}$ ($M=5, T=1600$): Ablation of abstention, hop-level fault isolation, and shadow probing ($\mathrm{Mean} \pm 1.96\,\mathrm{SEM}$; best per column in \textbf{bold}).}
\label{tab:tier1_e2}
{\small\setlength{\tabcolsep}{5pt}
\begin{tabular}{lccc}
\toprule
Variant & Tail Completion (\%) & Innocent Mis-Penalty (\%) & Toxic $\le 5\%$ (eps) \\
\midrule
\textbf{SYLVA Complete} & $\mathbf{83.4 \pm 3.5}$ & $2.00 \pm 0.18$ & $\mathbf{172.0 \pm 89.9}$ \\
A1 No Abstain Protection & $77.6 \pm 4.9$ & $\mathbf{1.82 \pm 0.12}$ & $492.0 \pm 104.9$ \\
A2 Uniform Chain Blame & $13.7 \pm 7.7$ & $21.67 \pm 0.72$ & $260.0 \pm 21.5$ \\
A3 No Shadow Probe & $73.8 \pm 5.3$ & $2.02 \pm 0.15$ & $384.0 \pm 267.1$ \\
\bottomrule
\end{tabular}}
\end{table}

\paragraph{Panel (c): E3---Resisting copycat wrappers and Sybil floods.}
Figure~\ref{fig:tier1}(c) plots the share of total market revenue reaching original core creators before and after a sudden flood at $t=600$ (vertical dashed line), when we inject $120$ copycat wrappers (half lineage-registered with $+0.004$ markup, half unlinked proxies with $+0.008$ markup, all seeded with $4.9$-star ratings) plus $12$ colluding Sybils. Naive rating leaderboards are instantly hijacked and fall to $\mathbf{0.006\% \pm 0.012\%}$ creator revenue share and genuine-provider routing share. With SYLVA ($\lambda_f = 18.0$ fee penalty, $85\%$ creator royalties, and a $0.002$-credit wash-settlement burn [$0.2\times f_v$] per fake self-call), original specialists hold steady at $\mathbf{86.2\% \pm 8.4\%}$ revenue share and $\mathbf{76.9\% \pm 15.0\%}$ genuine-provider routing share (logged as \texttt{snr} in code), while Sybil rings lose $\mathbf{2.513 \pm 0.046}$ currency units per 100 episodes (Table~\ref{tab:tier1_e3e4e5}).

\paragraph{Panel (d): E4---Recovering from silent rule drift.}
Figure~\ref{fig:tier1}(d) tracks completion across an unannounced statutory change at $t=800$ (vertical dashed line) that flips rules in half of all jurisdictions. To isolate drift adaptation cleanly without toxic confounders, $\mathrm{E4}$ evaluates clean specialist and generalist pools (omitting toxic hallucinators and careless planners, and directing $75\%$ of post-drift traffic to mutated jurisdictions via even-episode prioritization). At $t=800$, stale incumbents cause completion to drop sharply; SYLVA activates discounted forgetting ($\gamma = 0.988$) and a $6\%$ rookie (\texttt{canary}) quota, rebounding past $\ge 80\%$ in $\mathbf{280.0 \pm 24.8}$ episodes ($88.0\% \pm 2.9\%$ final completion; $67.6\% \pm 6.5\%$ patched-specialist win rate), whereas a global leaderboard fails to recover by $t=1600$ ($>800$ episodes, $30.6\% \pm 7.0\%$ final completion).

\paragraph{E5---Compiling 4-step chains into 1-call skills.}
Under stationary rules ($\gamma = 1.0$), skill compilation in $\mathrm{E5}$ ($N_{\min}=18$, $p_{\min}=0.90$) converts $3.8 \pm 1.9$ of the $8$ jurisdictions into 1-hop fast skills, slashing per-task tokens by $\mathbf{35.3\%}$ ($2673.8$ vs.\ $4132.0$) and executed hops by $\mathbf{34.4\%}$ ($2.62$ vs.\ $4.00$) at equivalent completion ($76.8\%$ vs.\ $79.7\%$; Table~\ref{tab:tier1_e3e4e5}).

\begin{table}[!t]
\centering
\caption{Tier-1 $\mathrm{E3}$--$\mathrm{E5}$ ($M=5, T=1600$): Wrapper/Sybil resistance, drift adaptation, and skill compilation ($\mathrm{Mean} \pm 1.96\,\mathrm{SEM}$).}
\label{tab:tier1_e3e4e5}
{\footnotesize\setlength{\tabcolsep}{5pt}
\begin{tabular}{llcc}
\toprule
Exp. & Metric & \textbf{SYLVA} & Baseline / Control \\
\midrule
$\mathrm{E3}$ & Core Revenue Share Post-Flood (\%) & $\mathbf{86.2 \pm 8.4}$ & $0.006 \pm 0.012$ (Naive) \\
$\mathrm{E3}$ & Genuine-Provider Routing Share (\texttt{snr}, \%) & $\mathbf{76.9 \pm 15.0}$ & $0.006 \pm 0.012$ (Naive) \\
$\mathrm{E3}$ & Sybil Capital Drain (units / 100 eps) & $\mathbf{2.513 \pm 0.046}$ & $0$ (zero-cost wash) \\
\midrule
$\mathrm{E4}$ & Post-Drift Recovery Time (eps to $\ge 80\%$) & $\mathbf{280.0 \pm 24.8}$ & $>800$ (Global, unrecovered) \\
$\mathrm{E4}$ & Post-Drift Tail Completion (\%) & $\mathbf{88.0 \pm 2.9}$ & $30.6 \pm 7.0$ (Global) \\
$\mathrm{E4}$ & Patched Specialist Win Rate (\%) & $\mathbf{67.6 \pm 6.5}$ & $47.9 \pm 15.2$ (Global) \\
\midrule
$\mathrm{E5}$ & Protocol Tokens / Task & $\mathbf{2673.8 \pm 697.5}$ (\textbf{-35.3\%}) & $4132.0 \pm 31.0$ (No Skill Compile) \\
$\mathrm{E5}$ & Executed Hops / Task & $\mathbf{2.62 \pm 0.67}$ (\textbf{-34.4\%}) & $4.00 \pm 0.00$ (No Skill Compile) \\
\bottomrule
\end{tabular}}
\end{table}

\section{Experiments II: Live Frozen-LLM Ecosystem (Tier-2)}

\subsection{Live LLM Setup and API Usage}

Tier-2 evaluates SYLVA under realistic LLM extraction noise and JSON formatting variance. We replace Tier-1 rule stubs with live frozen calls invoking \texttt{deepseek-chat} on DeepSeek's OpenAI-compatible API---whose HTTP response metadata resolves the serving snapshot as \texttt{deepseek-flash} (a single frozen backbone used throughout, with temperature $0.0$, max tokens $360$, and JSON Mode). Each task renders a messy commercial memo containing distractor credit limits and \texttt{Net-30} payment-term clauses, evaluated within strict $\tau = \$0.05$ tolerance. Tier-2 employs a compact 32-agent pool per suite ($8$ nodes per stage: $4$ regional specialists---each covering $2$ of the $K=8$ jurisdictions---alongside $2$ generalists prompted with non-localized default rates and $2$ prompt-injected hallucinators) across $M=3$ seeds (\texttt{[42, 101, 202]}) and $T=320$ episodes (tail window $t \in [260, 320)$) over a deterministic 24-invoice pool backed by an exact-match SQLite prompt cache.

\paragraph{Mapping Tier-1 to Tier-2 Suites.}
Tier-2 evaluates the four LLM-sensitive mechanisms, retaining our code/figure Phase-2 labels \texttt{P2-E1}--\texttt{P2-E4}: \texttt{P2-E1} (learning curve, matching $\mathrm{E1}$), \texttt{P2-E2} (fault-isolation ablation, matching $\mathrm{E2}$), \texttt{P2-E3} (drift recovery with simulated plain-English patches, corresponding to $\mathrm{E4}$), and \texttt{P2-E4} (single-call LLM skill compilation, corresponding to $\mathrm{E5}$); see Table~\ref{tab:tier2_summary}. We omit repeating $\mathrm{E3}$ in Tier-2 because pass-through wrappers invoke identical frozen downstream completions, making fee drainage a settlement and routing accounting property already demonstrated in Section~5 (and enforced on-chain in Section~\ref{sec:tier3_starknet}). Reported intervals give $\mathrm{Mean} \pm 1.96\,\mathrm{SEM}$ across seeds (normal approximation across $M=3$, clipped to $[0, 100\%]$ for percentages; e.g., $98.3 \pm 3.3\%$ yields $[95.1, 100.0]\%$).

\paragraph{Verified Live API Execution.}
Across all four suites ($\mathrm{P2\text{-}E1}$--$\mathrm{P2\text{-}E4}$), execution logged $\mathbf{6,543}$ live HTTPS API completions and $\mathbf{44,342}$ deterministic cache hits with $\mathbf{0}$ failures (average live latency $\mathbf{1,357.1\,\mathrm{ms}}$), consuming $\mathbf{4,210,801}$ live API tokens ($3,476,847$ prompt $+$ $733,954$ completion; $33,087,644$ effective simulation tokens) in $1,684.9\,\mathrm{s}$ total runtime (including $159.71\,\mathrm{s}$ prewarm).

\subsection{Quantitative Frozen-LLM Results}

\begin{figure}[!t]
\centering
\includegraphics[width=\textwidth]{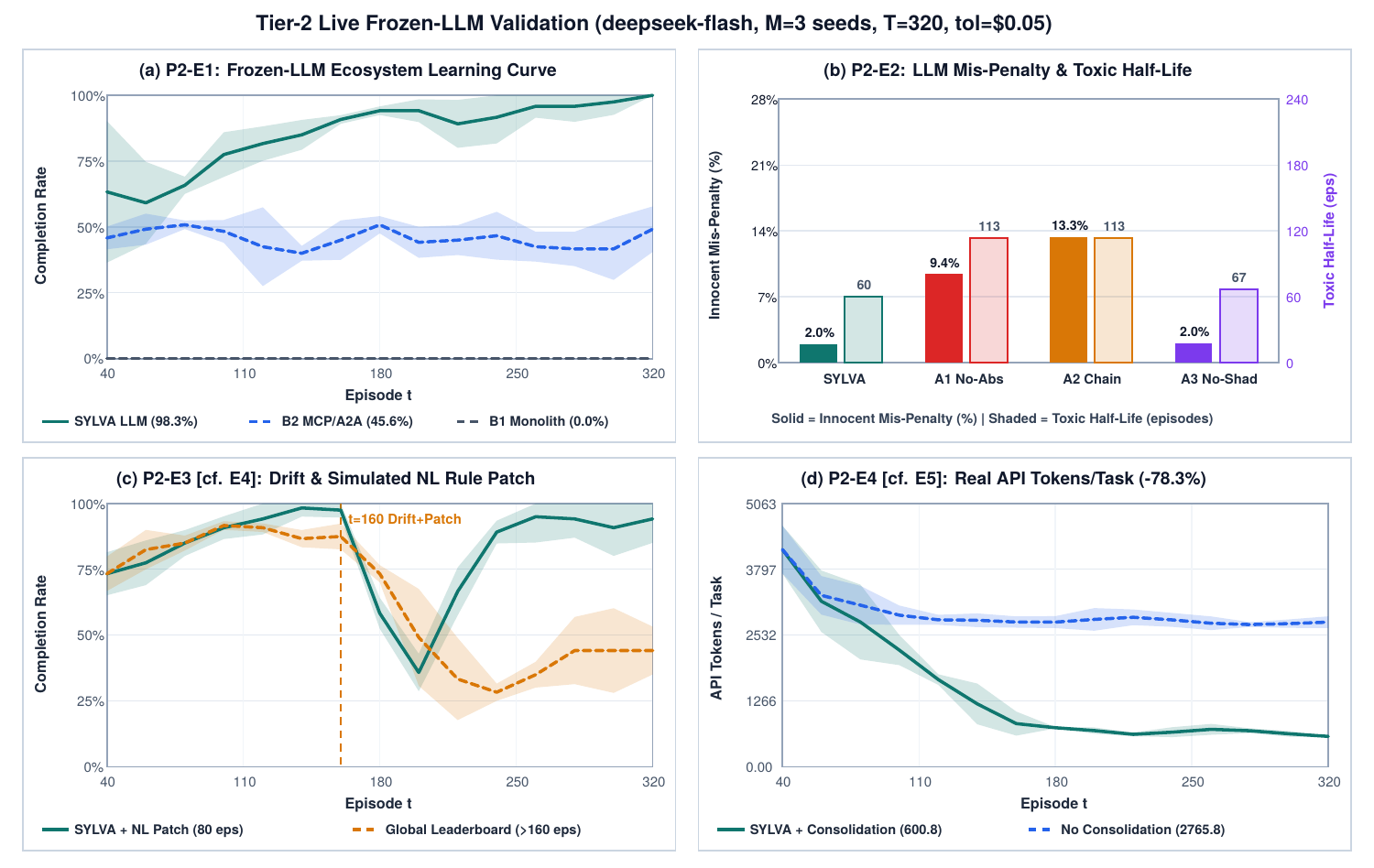}
\caption{Tier-2 live frozen-LLM results (\texttt{deepseek-flash}, $M=3$ seeds, $T=320$ episodes, $\tau=\$0.05$; line plots show $40$-episode rolling averages starting at $t=40$, and shaded bands show $\mathrm{Mean} \pm 1.96\,\mathrm{SEM}$ clipped to $[0, 100\%]$): \textbf{(a) P2-E1 Learning Curve}---end-to-end completion vs.\ episode $t$, where SYLVA reaches $98.3\%$ while static MCP/A2A ($\mathrm{B2}$) plateaus at $45.6\%$ and a standalone frozen LLM ($\mathrm{B1}$, dashed line flat along $0.0\%$) fails every invoice; \textbf{(b) P2-E2 Attribution \& Abstention Ablation}---dual-axis comparison of false penalties on innocent specialists (left axis, solid bars; lower is better) and the earliest $40$-episode window suppressing toxic traffic to $\le 5\%$ (right axis, shaded bars, labeled \texttt{Toxic Half-Life}; lower is better); \textbf{(c) P2-E3 Drift \& Plain-English Rule Patch}---completion across an unannounced statutory shift at $t=160$ paired with plain-English rule patches (labeled \texttt{NL Patch} in legend; vertical dashed line), where SYLVA restores $\ge 80\%$ in $80$ episodes vs.\ $>160$ unrecovered for a global leaderboard; \textbf{(d) P2-E4 Real API Token Savings}---measured DeepSeek API tokens per task, where 1-call skill compilation (labeled \texttt{Consolidation} in legend) saves $78.3\%$ of API tokens ($600.8$ vs.\ $2,765.8$).}
\label{fig:tier2}
\end{figure}

\begin{table}[!t]
\centering
\caption{Tier-2 Frozen-LLM Benchmarks (\texttt{P2-E1}, \texttt{P2-E3}, \texttt{P2-E4}; live \texttt{deepseek-flash}, $M=3, T=320, \tau=\$0.05$, $\mathrm{Mean} \pm 1.96\,\mathrm{SEM}$).}
\label{tab:tier2_summary}
{\footnotesize\setlength{\tabcolsep}{4pt}
\begin{tabularx}{\textwidth}{l>{\raggedright\arraybackslash}Xcc}
\toprule
Exp. & Metric & \textbf{SYLVA (Frozen LLM)} & Baseline \\
\midrule
$\mathrm{P2\text{-}E1}$ & Tail Completion Rate (\%) & $\mathbf{98.3 \pm 3.3}$ ($[95.1, 100.0]$) & $45.6 \pm 6.1$ ($\mathrm{B2}$), $0.0 \pm 0.0$ ($\mathrm{B1}$) \\
\midrule
$\mathrm{P2\text{-}E3}$ [cf.\ $\mathrm{E4}$] & Recovery Time Post-Drift (eps to $\ge 80\%$, 20-ep grid) & $\mathbf{80.0 \pm 0.0}$ & $>160$ (Global, unrecovered) \\
$\mathrm{P2\text{-}E3}$ [cf.\ $\mathrm{E4}$] & Post-Drift Completion (\%) & $\mathbf{93.9 \pm 7.1}$ & $45.6 \pm 9.7$ (Global) \\
$\mathrm{P2\text{-}E3}$ [cf.\ $\mathrm{E4}$] & Patched Specialist Win Rate (across all 320 eps, \%) & $\mathbf{56.2 \pm 6.5}$ & $47.2 \pm 2.8$ (Global) \\
\midrule
$\mathrm{P2\text{-}E4}$ [cf.\ $\mathrm{E5}$] & Real API Tokens / Task & $\mathbf{600.8 \pm 36.3}$ (\textbf{-78.3\%}) & $2,765.8 \pm 85.6$ (No Skill Compile) \\
$\mathrm{P2\text{-}E4}$ [cf.\ $\mathrm{E5}$] & Executed Hops / Task & $\mathbf{1.10 \pm 0.06}$ (\textbf{-72.5\%}) & $4.00 \pm 0.00$ (No Skill Compile) \\
$\mathrm{P2\text{-}E4}$ [cf.\ $\mathrm{E5}$] & Tail Completion / Compiled Jurisdictions & $\mathbf{96.1 \pm 2.9}\%$ ($8.0/8$) & $94.4 \pm 4.7\%$ ($0/8$) \\
\bottomrule
\end{tabularx}}
\end{table}

\begin{table}[!t]
\centering
\caption{Tier-2 $\mathrm{P2\text{-}E2}$ ($M=3, T=320$): Live frozen-LLM attribution \& abstention ablation ($\mathrm{Mean} \pm 1.96\,\mathrm{SEM}$; best per column in \textbf{bold}).}
\label{tab:tier2_e2}
{\footnotesize\setlength{\tabcolsep}{5pt}
\begin{tabular}{lccc}
\toprule
Variant & Tail Completion (\%) & Innocent Mis-Penalty (\%) & Toxic $\le 5\%$ (eps) \\
\midrule
\textbf{SYLVA Complete} & $96.7 \pm 3.8$ & $\mathbf{1.96 \pm 0.83}$ & $\mathbf{60.0 \pm 39.2}$ \\
A1 No Abstain Protection & $\mathbf{98.3 \pm 1.9}$ & $9.39 \pm 0.81$ ($4.78\times$) & $113.3 \pm 13.1$ ($1.89\times$) \\
A2 Uniform Chain Blame & $96.1 \pm 1.1$ & $13.33 \pm 3.39$ ($6.79\times$) & $113.3 \pm 13.1$ ($1.89\times$) \\
A3 No Random Shadow Probe & $96.7 \pm 3.8$ & $2.01 \pm 0.73$ & $66.7 \pm 52.3$ ($1.11\times$) \\
\bottomrule
\end{tabular}}
\end{table}

Figure~\ref{fig:tier2} and Tables~\ref{tab:tier2_summary}--\ref{tab:tier2_e2} demonstrate every core mechanism on live frozen LLM completions:
\begin{itemize}
    \item \textbf{Beating a Single Frozen LLM and Static Routing ($\mathrm{P2\text{-}E1}$)}: Lacking private regional rules, the standalone monolithic LLM ($\mathrm{B1}$) scores $\mathbf{0.0\% \pm 0.0\%}$ within $\$0.05$ tolerance, and static MCP/A2A routing ($\mathrm{B2}$) plateaus at $45.6\% \pm 6.1\%$. SYLVA reaches $\mathbf{98.3\% \pm 3.3\%}$ with zero weight updates.
    \item \textbf{Why Whole-Chain Blame Hurts Worse in Larger Markets ($\mathrm{P2\text{-}E2}$)}: Removing safe abstention ($\mathrm{A1}$) inflates false specialist penalties by $\mathbf{4.78\times}$ ($9.39\%$ vs.\ $1.96\%$) and slows toxic clearance by $\mathbf{1.89\times}$ ($113.3$ vs.\ $60.0$ episodes). Converged tail completion in Table~\ref{tab:tier2_e2} still overlaps ($98.3\% \pm 1.9\%$ vs.\ $96.7\% \pm 3.8\%$) because traffic at $t \in [260, 320)$ concentrates on matched in-domain specialists that rarely hit OOD boundaries, so the real penalty of dropping abstention protection shows up in unnecessary specialist blame and lingering toxic traffic during exploration. Removing step-level blame isolation ($\mathrm{A2}$) inflates false penalties by $\mathbf{6.79\times}$ ($13.33\%$) and slows clearance by $\mathbf{1.89\times}$. Comparing $\mathrm{E2}$ and $\mathrm{P2\text{-}E2}$ highlights a striking market-size effect: because Tier-2's generalists use fixed inaccurate default rates (scoring $0\%$ within $\pm\$0.05$) and hallucinators nearly always fail, compliant specialists still retain traffic despite absorbing $13.33\%$ false blame ($96.1\%$ completion in Table~\ref{tab:tier2_e2}). In Tier-1's larger 64-agent pool with $72\%$-accurate generalists, by contrast, $21.67\%$ false penalties push specialists below generalists and collapse completion to $13.7\%$. Whole-chain blame thus looks harmless in small sparse pools yet collapses once competitors are stronger. Meanwhile, keeping disputed 3-peer replay active when disabling random shadow probes ($\mathrm{A3}$) slows toxic clearance modestly ($66.7$ vs.\ $60.0$ eps, $1.11\times$).
    \item \textbf{Plain-English Rule Patching After Drift ($\mathrm{P2\text{-}E3}$)}: Mirroring $\mathrm{E4}$ cleanly without toxic confounders, stale incumbents in drifted jurisdictions are supplemented at $t=160$ by fresh cold-start specialists patched in plain English (with $75\%$ post-drift traffic directed to drifted jurisdictions). Evaluated on a 20-episode sliding grid, SYLVA restores $\ge 80\%$ completion in $\mathbf{80.0 \pm 0.0}$ episodes across all three seeds ($93.9\% \pm 7.1\%$ post-drift completion; patched specialists win $\mathbf{56.2\% \pm 6.5\%}$ of all $320$ episodes as un-drifted incumbents continue serving unchanged jurisdictions), outperforming an unrecovered global leaderboard ($>160$ eps, $45.6\% \pm 9.7\%$).
    \item \textbf{Single-Call Skill Compilation ($\mathrm{P2\text{-}E4}$)}: Compiling stable 4-hop chains strictly from participating rulebooks ($N_{\min}=8, p_{\min}=0.88$) converts all $\mathbf{8.0 / 8}$ jurisdictions into single-call skills, slashing real API tokens by $\mathbf{78.3\%}$ ($600.8$ vs.\ $2,765.8$ tokens/task) and hops by $\mathbf{72.5\%}$ ($1.10$ vs.\ $4.00$).
\end{itemize}

\begin{table}[!t]
\centering
\caption{Hyperparameters across Tier-1 (structural simulation) and Tier-2 (frozen LLM). Tier-2 forgetting $\gamma{=}0.96$ and threshold $N_{\min}{=}8$ scale proportionally with its $5\times$ shorter horizon ($T{=}320$ vs.\ $1600$); fee weight $\lambda_f{=}18.0$ targets $\mathrm{E3}$ wrapper markups (equal $f_v{=}0.010$ makes $\lambda_f{=}0$ identical elsewhere). Live Starknet Sepolia (Tier-3) uses $f_v{=}0.02\,\mathrm{STRK}$, $0.04\,\mathrm{STRK}$ wrappers/skills, $0.001\,\mathrm{STRK}$ OOD fee ($5\%$), $2.0\,\mathrm{STRK}$ stake, and $0.50\,\mathrm{STRK}$ slash.}
\label{tab:hyperparams}
{\footnotesize\setlength{\tabcolsep}{3.5pt}
\begin{tabularx}{\textwidth}{l>{\raggedright\arraybackslash}Xcc}
\toprule
Component & Parameter & Tier-1 Value & Tier-2 Value \\
\midrule
Environment & Contexts $K$ / Stages $H$ / Dim $d$ & $8$ / $4$ / $12$ & $8$ / $4$ / $12$ \\
Environment & Episodes $T$ / Seeds $M$ / Tol $\tau$ & $1600$ / $5$ / $10^{-3}$ & $320$ / $3$ / $\$0.05$ \\
Planners & Fault prob.\ $q$ (benign / faulty) & $0.02$ / $0.35$ & $0.02$ / $0.35$ \\
Economics & Base step fee $f_v$ / Initial stake & $0.010$ / $1.0$ credits & $0.010$ / $1.0$ credits \\
Clearinghouse & Wrapper share $\rho_{\mathrm{core}}$ / OOD fee / Slash & $0.85$ / $0.0005$ / $0.08$ & $0.85$ / $0.0005$ / $0.08$ \\
Boundary Check & Assertion noise $(\beta_{\mathrm{fn}}, \beta_{\mathrm{tox}}, \alpha_{\mathrm{fp}})$ & $(0.12, 0.60, 0.02)$ & $(0.12, 0.60, 0.02)$ \\
Shadow Probe & Shadow probe rate ($\mathrm{E1}$, $\mathrm{E3}$--$\mathrm{E5}$ / $\mathrm{E2}$) & $0.05$ / $0.08$ & $0.08$--$0.10$ \\
Bandit Router & Ridge $\lambda$ / $\kappa$ / Forgetting $\gamma$ & $1.0$ / $0.40$--$0.45$ / $1.0$ ($0.988$) & $1.0$ / $0.40$--$0.45$ / $1.0$ ($0.96$) \\
Fees / Canary & Fee penalty $\lambda_f$ ($\mathrm{E3}$ / other) / Canary & $18.0$ / $0.0$ / $0.03$--$0.06$ & $0.0$ / $0.0$ / $0.05$--$0.10$ \\
Skill Compiler & Consolidator $(N_{\min}, p_{\min})$ & $(18, 0.90)$ & $(8, 0.88)$ \\
\bottomrule
\end{tabularx}}
\end{table}

\subsection{Case Study: Two Real Episodes on the Interactive Console}

To walk through one complete run while keeping outputs deterministic, the open CLI console (\texttt{poc/interactive\_symbiont.py --demo}) runs live \texttt{deepseek-flash} calls with real RFC~8032 Ed25519 signing and synchronous SQLite commits ($41$ live HTTPS calls $+$ $300$ cache hits, $26,146$ real API tokens, $0$ failures). Across its verified 24-episode audit trace ($241$ signed Ed25519 receipts, $27$ royalty payouts, and $27$ stake slashes), two short episodes illustrate SYLVA step-by-step:
\begin{enumerate}
    \item \textbf{Supply-Side Plain-English Rule Update (\texttt{US-SaaS-Tax})}: Pre-drift exploration completes $10/24$ ($41.7\%$) episodes ($2/10$ on \texttt{US-SaaS-Tax}). After \texttt{US-SaaS-Tax} shifts silently (true tax $\$116.6815$), an unpatched LLM specialist outputs $\$58.1731$ ($\$58.5084$ error). Injecting a plain-English rule patch lowers Stage-1 and Stage-2 errors to $\$0.0000$ ($\$116.6815$ and $\$43.5185$ exact). Across the next $24$ \texttt{US-SaaS-Tax} episodes, completion rises to $17/24$ ($70.8\%$) overall and $\mathbf{8/10}$ ($\mathbf{80.0\%}$) in the tail, while the core tax specialist collects $85\%$ creator royalties from its registered wrapper.
    \item \textbf{Demand-Side Stealth Hallucination Isolation (\texttt{EU-MedDevice})}: When a Stage-1 hallucinator fabricates \texttt{tax\_fee} as $\$115.3788$ (vs.\ truth $\$134.9817$) inside valid JSON, end-user rejection triggers 3-peer step replay (an accounting estimate of $2,600$ protocol replay tokens across $13$ cumulative stage probes in the 24-episode trace at $200$ estimated tokens per probe, separate from the $26,146$ measured DeepSeek API tokens reported above). Replay isolates the Stage-1 hallucinator against a Stage-1 specialist, slashing the hallucinator's bonded stake ($-0.6$ utility penalty) while shielding all three innocent specialists ($0$ false penalties, $+0.25$ protection credit each).
\end{enumerate}

\section{Experiments III: Production Realization on Starknet \& Rust \texttt{rmcp} (Tier-3)}
\label{sec:tier3_starknet}

To bridge controlled multi-seed simulation with real-world deployment and remove all Tier-1/Tier-2 POC oracle abstractions, we implemented the complete protocol in domain-agnostic Rust (\texttt{rmcp 3.11} SDK, Starknet v3 client, \texttt{sylva-cli}) paired with Cairo~2 smart contracts (\texttt{SylvaClearinghouse v0.2.0} and \texttt{verifiable\_router}) settled in Starknet's native \texttt{STRK} token \cite{bensasson2018stark,goldberg2021cairo}.

\subsection{Domain-Agnostic Paid MCP Architecture \& Production Safeguards}

Rather than placing heavy semantic routing or inference on-chain, the Tier-3 stack enforces a strict boundary between fast off-chain MCP negotiation and minimal cryptographic on-chain settlement:
\begin{enumerate}
    \item \textbf{Off-Chain Router Gateway, Two-Tier Decentralized Memory \& Rust \texttt{rmcp} SDK}: The router gateway (\texttt{sylva-router-mcp}) runs completely off-chain---either locally on the caller's client as a zero-trust MCP sidecar or as an open gateway with zero custody over funds. Authoritative node stakes, caller escrow, active compiled skills, and the neutral \texttt{Q32.32} audit verifier reside on Starknet. This two-tier memory design avoids prohibitive per-hop on-chain storage ($d^2=144$ fixed-point entries per hop) and centralized routing lock-in. First, registered 1-call \texttt{CompiledSkill} routes, node \texttt{active} status, and settlement/slash events (\texttt{HopSettled}, \texttt{NodeSlashed}) live permanently on Starknet, so any fresh gateway can bootstrap read-only priors and suspensions directly via public RPC events (\texttt{sync-router-state}) without private keys. Second, fine-grained System-2 matrices $(A_{t,v}, b_{t,v})$ checkpoint atomically off-chain (\texttt{--state-file}) so private caller contexts $x_t$ remain local. On the supply side, any provider implements the three-method \texttt{SylvaStageHandler} trait over \texttt{stdio} or remote \texttt{TCP}. It handles arbitrary JSON schemas via key-order-invariant Poseidon (STARK-friendly hash) digests, projects domain strings via signed feature hashing to $\mathbb{R}^{12}$, supports runtime \texttt{patch\_rulebook} updates and \texttt{OOD-ABSTAIN} signaling, and signs every handoff with a 9-element Poseidon domain digest (\texttt{'SYLVA\_MCP\_V1'}) under its StarkCurve key.
    \item \textbf{Zero-Oracle Quorum Slashing \& Low-Stake Deactivation}: Disputes never consult hidden ground truth. When a caller rejects an episode ($Y_t = 0$), \texttt{resolve\_dispute\_quorum} verifies the accused node's signature and at least $m \ge 3$ independent, currently staked peer replays matching the disputed \texttt{(episode\_id, hop, input\_hash)}. Only when a $\ge 2/3$ supermajority agrees on an \texttt{output\_hash} conflicting with the accused node does the contract slash $L_{\mathrm{slash}} = 0.50\,\mathrm{STRK}$ ($25\times$ the $0.02\,\mathrm{STRK}$ standard hop fee) into \texttt{caller\_escrow}. Because innocent steps and the $m=3$ peer-replay receipts are still settled on rejected runs ($Y_t = 0$), this slash reimburses both the faulty hop and peer-replay fees while deterring frivolous disputes, and automatically sets \texttt{active = false} whenever remaining stake drops below $L_{\mathrm{slash}}$.
    \item \textbf{On-Chain OOD Cap, Lineage Splits, Burn, and Skill Circuit Breaker}: After a 3-call warmup, \texttt{SylvaClearinghouse} enforces a hard $40\%$ abstention cap (\texttt{max\_abstain\_bps = 4000}, zero micro-fee above cap), routes configurable $85\% / 15\%$ wrapper creator royalties and $4 \times 25\%$ compiled-skill splits through a pull-payment ledger after deducting a $1\%$ (\texttt{100 bps}) non-refundable burn, and disables any compiled skill whose sliding-window pass rate drops below $80\%$ over $\ge 5$ calls.
    \item \textbf{Progressive Governance, Never-Paused Exit \& Micro-Episode Solvency}: Version \texttt{0.2.0} supports native \texttt{replace\_class\_syscall} upgrades guarded by an $86,400\,\mathrm{s}$ (24-hour) public timelock, a one-way delay ratchet (\texttt{increase\_upgrade\_delay}), and two-step ownership handover before permanently locking contract code against future upgrades (\texttt{ossify()}). Crucially, unconditional exit entrypoints (\texttt{deactivate\_and\_unstake}, \texttt{withdraw\_escrow}, \texttt{claim\_earnings}) bypass circuit-breaker pauses so participants can withdraw 100\% of funds before any upgrade activates. Because per-episode fees are micro-payments ($\sim 0.08\,\mathrm{STRK}$ across four hops), pre-funded escrow proves caller solvency upfront, and gateways or providers immediately drop any caller address that withholds post-episode settlement (bounding exposure to one micro-episode, while timelocked per-episode holds remain straightforward for high-value calls).
    \item \textbf{Bit-Exact Q32.32 Verifiable Router \& Poseidon State Commitments}: Off-chain Disjoint LinUCB scoring (\texttt{score\_arm\_q32}) and on-chain \texttt{verify\_linucb\_score\_q32} share identical toward-zero \texttt{i128} fixed-point arithmetic ($2^{32}$ scale, integer square root), permitting permissionless on-chain verification that gateways select arms neutrally by Eq.~\eqref{eq:ucb}. In addition, each off-chain checkpoint computes a lexicographically sorted Poseidon commitment (\texttt{'SYLVA\_RTR\_V1'}) over all arms' \texttt{Q32.32} $(\hat{\theta}_{q32}, A_{q32}^{-1})$ parameters (exposed via \texttt{get\_router\_state\_summary}) to anchor open-gateway audits.
\end{enumerate}

\subsection{Live Starknet Sepolia E2E \& Contract Verification}

The Cairo contracts pass $12/12$ Starknet Foundry (\texttt{snforge}) unit and integration tests (covering STARK signatures, 3-peer quorum splitting, OOD caps, skill circuit breakers, upgrade preservation, pause-bypassing exits, and \texttt{ossify()}), while the 5-crate Rust workspace passes $9/9$ unit, persistence, and multi-process TCP \texttt{rmcp} tests. Local \texttt{starknet-devnet} runs further exercise a live \texttt{replace\_class\_syscall} upgrade from \texttt{0.2.0} to \texttt{0.3.0-upgraded} under preserved state, alongside automated cold-start event sync and snapshot reload. On Starknet Sepolia (\texttt{drpc.org}), we deployed \texttt{SylvaClearinghouse v0.2.0} (contract \texttt{0x0676b73c...917fe12a}, class hash \texttt{0x04cf8563...0bed05}, $86,400\,\mathrm{s}$ timelock) and ran all Paid MCP market scenarios across $19$ live Starknet v3 transactions ($1$ declare, $1$ deploy, $9$ node registrations, and $8$ scenario transactions) using $2.0\,\mathrm{STRK}$ stake/node, $0.50\,\mathrm{STRK}$ slash, $5.0\,\mathrm{STRK}$ escrow, $0.02\,\mathrm{STRK}$ standard hop fees, $0.04\,\mathrm{STRK}$ wrapper/skill fees, and $0.001\,\mathrm{STRK}$ ($5\%$) OOD micro-fees. Independent read-only checks via Starknet Foundry's \texttt{sncast call} and zero-key \texttt{sync-router-state} verified all balances, bit-exact Q32.32 scores, all $9$ nodes, $21$ emitted events, \texttt{tax\_p3\_s1} suspension, \texttt{hallu\_s1} slash, active \texttt{skill\_eu0}, and deterministic Poseidon commitment \texttt{0x03b86cca...c3ecb7} (Table~\ref{tab:tier3_sepolia}; full 66-character hashes in \texttt{results/starknet\_sepolia\_e2e\_summary.json}).

\begin{table}[!t]
\centering
\caption{Tier-3 live Starknet Sepolia \texttt{v0.2.0} E2E verification (\texttt{0x0676b73c...917fe12a}, \texttt{drpc.org}, $19$ native \texttt{STRK} transactions).}
\label{tab:tier3_sepolia}
{\footnotesize\setlength{\tabcolsep}{4pt}
\begin{tabularx}{\textwidth}{>{\raggedright\arraybackslash}l>{\raggedright\arraybackslash}X>{\raggedright\arraybackslash}l}
\toprule
Stage / Check & On-Chain Action & Verified State / Outcome (\texttt{sncast call}) \\
\midrule
Setup (11 txs) & Declare + deploy \texttt{v0.2.0} + register $9$ nodes & \texttt{version="0.2.0"}, \texttt{delay=86400s}, $2.0\,\mathrm{STRK}$/node \\
Scen.\ 1: Deposit (1 tx) & Caller locks escrow (\texttt{0x068c82b2...d2b2fd}) & $5.0\,\mathrm{STRK}$ ($5\times 10^{18}\,\mathrm{fri}$; $1\,\mathrm{STRK}{=}10^{18}\,\mathrm{fri}$) \\
Scen.\ 2: OOD (1 tx) & \texttt{tax\_jp\_s1} abstains (\texttt{0x033e345d...8ecd167}) & \texttt{abstain\_calls=1}, \texttt{stake=2.000000 STRK} (zero slash) \\
Scen.\ 3: Wrapper (2 txs) & $85/15$ + \texttt{claim\_earnings} (\texttt{0x0237333a...a2f95a4}) & Core royalty $0.033660$, wrapper $0.005940\,\mathrm{STRK}$ \\
Scen.\ 4: Quorum (1 tx) & $m{=}3$ peers slash \texttt{hallu\_s1} (\texttt{0x00a47496...a0acf8}) & \texttt{hallu\_s1} cut to $1.5\,\mathrm{STRK}$; $+0.5\,\mathrm{STRK}$ to escrow \\
Scen.\ 5: Skill (2 txs) & Register + 1-call $4{\times}25\%$ (\texttt{0x029c93f9...be3d36a2}) & \texttt{skill\_eu0} wins $=1$; cumul.\ $1\%$ burn $0.002210\,\mathrm{STRK}$ \\
Scen.\ 6: Exit (1 tx) & \texttt{deactivate\_and\_unstake} (\texttt{0x0688feff...eeabe26}) & \texttt{tax\_p3\_s1} \texttt{active=false}, $2.0\,\mathrm{STRK}$ returned \\
Read-Only Score Audit & \texttt{verify\_linucb\_score\_q32} on-chain & Exact match: \texttt{rust\_q32 == cairo\_q32 == 2741039193} \\
Read-Only Event Sync & \texttt{sync-router-state} (public RPC events) & $21$ events, $9$ nodes; \texttt{tax\_p3\_s1} off; \texttt{0x03b86cca...c3ecb7} \\
\bottomrule
\end{tabularx}}
\end{table}

\section{Discussion, Limitations \& Planned Human-Subject Study}

\subsection{Where Open Agent Markets Beat Closed Models \& Next Engineering Steps}

Even as Tier-3 turns the compliance benchmark design into a general-purpose Paid MCP clearinghouse, three market settings naturally favor open agent ecosystems over closed all-in-one apps, alongside two remaining engineering steps:
\begin{enumerate}
    \item \textbf{Where Open Markets Beat Closed All-in-One Models}: Closed apps starve external protocols of feedback, and stronger frontier models will absorb generic thin wrappers. Three kinds of services, however, cannot easily be folded into one giant model: (i)~\emph{cross-company boundaries} (such as buyer--seller or hospital--insurer negotiations whose privacy rules and conflicting interests forbid merging data or prompts), (ii)~\emph{live private state and real-world actuators}, and (iii)~\emph{bonded liability providers} willing to back regulated outputs (tax, clearing, clinical triage) with slashable deposits.
    \item \textbf{Open Hardening Directions}: First, while Tier-3 checks individual \texttt{Q32.32} routing scores on-chain, full multi-episode STARK trajectory proofs (e.g., via StarkWare's Circle-STARK prover \texttt{Stwo}) and dual-orchestrator quorum replay for ambiguous natural-language instructions remain future engineering work. Second, stopping hidden off-chain reselling outside clearinghouse receipts requires core services to bind wholesale API discounts strictly to verified lineage tokens.
\end{enumerate}

\subsection{Preprint Scope, Limitations \& Planned Two-Arm Human Trial}
\label{sec:human_study}

As a fast-release preprint combining formal proofs, multi-seed simulations, and a live Starknet deployment, this work keeps four clear experimental boundaries prior to archival submission:
\begin{enumerate}
    \item \textbf{Controlled Multi-Seed Simulator Sweeps (Tiers~1--2)}: Although Tier-3 enforces zero-oracle $m \ge 3$ quorum consensus, $< L_{\mathrm{slash}}$ suspension, $40\%$ OOD caps, and $<80\%$ compiled-skill circuit breakers live on Starknet, our large sweeps in Tiers~1--2 still use sandbox stage checks in \texttt{poc/protocol.py}. They also evaluate hazards modularly---activating forgetting $\gamma < 1$ without toxic/careless agents in drift suites and setting $\lambda_f = 18.0$ in wrapper-flood runs---rather than running hundreds of live blockchain episodes under all hazards simultaneously. Furthermore, right after an unannounced rule shift, if a caller disputes a lone newly patched pioneer while $\ge 3$ non-abstaining peers still hold stale rules, zero-oracle majority voting lags until $\ge 2$ peers patch or abstain.
    \item \textbf{Task Complexity}: Our invoice benchmark spans a fixed 4-stage sequential pipeline rather than arbitrary directed acyclic graph (DAG) workflows, multi-step software synthesis, or mixed-model agent pools (beyond \texttt{deepseek-flash}).
    \item \textbf{Tier-2 Sample Pool \& Statistical Approximation}: Tier-2 evaluates $M=3$ random seeds over a deterministic 24-invoice prompt cache, reporting normal-approximation $1.96$-SEM bands (rather than wider $t_{0.975,2}$ intervals) and omitting baselines such as peer-majority voting or LLM-as-a-judge sampling.
    \item \textbf{Simulated vs.\ Real Human Feedback}: Both supply-side rule patches and demand-side rejection signals in Tiers~1--2 are deterministic proxies rather than noisy, delayed, or ambiguous inputs from real practitioners.
\end{enumerate}

\paragraph{Planned Two-Arm Human Study \& Archival Roadmap.}
For the journal/conference version, we will evaluate all hazards together on mixed-model coding and tool-use workflows backed by peer-quorum replay, and recruit human participants on \texttt{poc/interactive\_symbiont.py} across a two-arm trial: (1)~\textbf{Arm~A (Free-Form Supply Maintenance)}, measuring frozen-LLM rule adherence under unscripted, noisy practitioner patches; and (2)~\textbf{Arm~B (Blinded Invoice Auditing)}, testing step-level blame isolation against noisy human acceptance feedback.

\section{Conclusion}

We presented SYLVA, a market protocol showing how frozen foundation models and human-maintained services can form an adaptive, self-correcting agent market without updating weights ($\Delta W = 0$). Combining formal incentive guarantees across OOD abstention, step-level blame isolation, and wrapper elimination with three-tier validation---a 64-agent market simulation, $6,543$ live DeepSeek completions under $\pm\$0.05$ tolerance, and a Rust \texttt{rmcp} $+$ Cairo \texttt{SylvaClearinghouse v0.2.0} deployment on Starknet Sepolia---SYLVA achieves $82.1\%$--$98.3\%$ end-to-end completion while cutting false specialist penalties by $6.79\times$--$10.8\times$. It further retains $86.2\%$ creator revenue under wrapper and Sybil floods, recovers within $80.0$ episodes in Tier-2 ($280.0$ in Tier-1) after silent rule shifts, and saves up to $78.3\%$ of real API tokens via compiled 1-call skills. Finally, our open Rust/Cairo release verifies zero-oracle 3-peer quorum slashing, bit-exact \texttt{Q32.32} routing, and timelocked trustless provider exit live on-chain, providing an extensible testbed for mixed-model benchmarks and human-subject evaluation.

\bibliographystyle{plain}
\bibliography{references}

\end{document}